\documentclass[
reprint, superscriptaddress,
amsmath,amssymb,
aps
]{revtex4-2}

\usepackage{graphicx}
\usepackage{siunitx}  
\usepackage{amsmath}
\usepackage{tabularx}
\usepackage{braket}   
\usepackage{comment}     
\usepackage[dvipsnames]{xcolor}
\usepackage[normalem]{ulem}
\usepackage{multirow}
\usepackage{float}

\date{\today}

\begin{document}

\title{Low loss lumped-element inductors made from granular aluminum}

\author{Vishakha Gupta}

\email{These authors contributed equally; vishakha.gupta@yale.edu, patrick.winkel@yale.edu, neel.thakur@yale.edu}

\affiliation{Departments of Applied Physics and Physics, Yale University, New Haven, CT, USA}
\affiliation{Yale Quantum Institute, Yale University, New Haven, CT, USA}

\author{Patrick Winkel}

\email{These authors contributed equally; vishakha.gupta@yale.edu, patrick.winkel@yale.edu, neel.thakur@yale.edu}

\affiliation{Departments of Applied Physics and Physics, Yale University, New Haven, CT, USA}
\affiliation{Yale Quantum Institute, Yale University, New Haven, CT, USA}

\author{Neel Thakur}

\email{These authors contributed equally; vishakha.gupta@yale.edu, patrick.winkel@yale.edu, neel.thakur@yale.edu}
\affiliation{Departments of Applied Physics and Physics, Yale University, New Haven, CT, USA}
\affiliation{Yale Quantum Institute, Yale University, New Haven, CT, USA}

\author{Peter van Vlaanderen}
\affiliation{Departments of Applied Physics and Physics, Yale University, New Haven, CT, USA}
\affiliation{Yale Quantum Institute, Yale University, New Haven, CT, USA}

\author{Yanhao Wang}
\affiliation{Departments of Applied Physics and Physics, Yale University, New Haven, CT, USA}
\affiliation{Yale Quantum Institute, Yale University, New Haven, CT, USA}

\author{Suhas Ganjam}
\thanks{Present address: Google Quantum AI, Santa Barbara, CA}
\affiliation{Departments of Applied Physics and Physics, Yale University, New Haven, CT, USA}
\affiliation{Yale Quantum Institute, Yale University, New Haven, CT, USA}

\author{Luigi Frunzio}
\affiliation{Departments of Applied Physics and Physics, Yale University, New Haven, CT, USA}
\affiliation{Yale Quantum Institute, Yale University, New Haven, CT, USA}

\author{Robert J. Schoelkopf}
\email{robert.schoelkopf@yale.edu}
\affiliation{Departments of Applied Physics and Physics, Yale University, New Haven, CT, USA}
\affiliation{Yale Quantum Institute, Yale University, New Haven, CT, USA}

\begin{abstract}

Lumped-element inductors are an integral component in the circuit QED toolbox. However, it is challenging to build inductors that are simultaneously compact, linear and low-loss with standard approaches that either rely on the geometric inductance of superconducting thin films or on the kinetic inductance of Josephson junctions arrays. In this work, we overcome this challenge by utilizing the high kinetic inductance offered by superconducting granular aluminum (grAl). We demonstrate lumped-element inductors with a few nH of inductance that are up to $100$ times more compact than inductors built from pure aluminum (Al). To characterize the properties of these linear inductors, we first report on the performance of lumped-element resonators built entirely out of grAl with sheet inductances varying from $30-320$\,pH/sq and self-Kerr non-linearities of $0.2-20\,\mathrm{Hz/photon}$. Further, we demonstrate ex-situ integration of these grAl inductors into hybrid resonators with Al or tantalum (Ta) capacitor electrodes without increasing total internal losses. Interestingly, the measured internal quality factors systematically decrease with increasing room-temperature resistivity of the grAl film for all devices, indicating a trade-off between compactness and internal loss. For our lowest resistivity grAl films, we measure quality factors reaching $3.5 \times 10^6$ for the all-grAl devices and $4.5 \times 10^6$ for the hybrid grAl/Ta devices, similar to state-of-the-art quantum circuits. Our loss analysis suggests that the surface loss factor of grAl is similar to that of pure Al for our lowest resistivity films, while the increasing losses with resistivity could be explained by increasing conductor loss in the grAl film.

\end{abstract}
\maketitle

\section{Introduction}
\begin{figure}[t]
    \includegraphics[width =\linewidth]{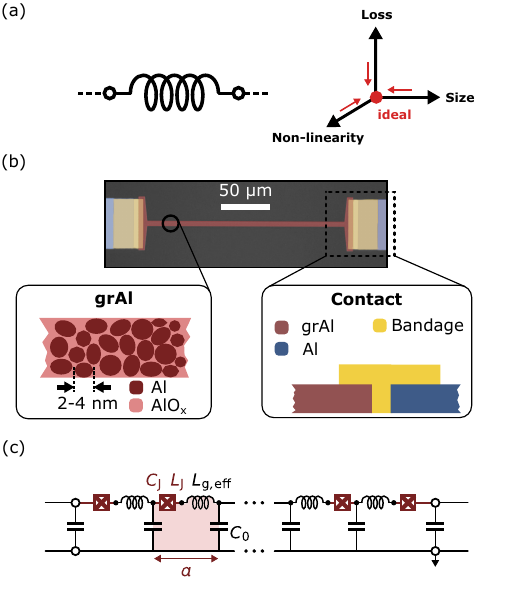}
    \caption{\textbf{Lumped-element inductor made from granular aluminum (grAl).} (a) Circuit representation of a lumped-element inductor. The ideal lumped-element inductor for superconducting quantum circuits simultaneously minimizes the footprint, non-linearity and loss at microwave frequencies. (b) False-color coded optical microscope image of a lumped-element inductor with $L_{\mathrm{k}} \approx 7.8\,$nH, realized with a thin-film of high-kinetic inductance grAl (red). The inductor is galvanically connected to an Al film (blue) through a bandage layer (yellow). In superconducting grAl, crystalline Al grains are separated by non-stoichiometric $\mathrm{AlO}_x$, resulting in a 3D array of JJs.  (c) Effective circuit model of the grAl inductor inspired by the granular microstructure. Due to the large aspect ratio, the grAl strip is modelled as a 1D array of effective JJs. Each effective junction of size $a$ contributes with a kinetic inductance $L_{\mathrm{J}}$ and a capacitance $C_{\mathrm{J}}$, alongside with a small geometric inductance $L_{\mathrm{g,eff}}$ and a capacitance to ground $C_0$. }
    \label{Fig1}
\end{figure}

Superconducting architectures for quantum information processing are growing increasingly complex\,\cite{Arute2019,Chen2021,Jurcevic_2021,Bravyi_outlook_2022,Xu_2023,acharya2024quantumerrorcorrectionsurface}, as individual circuit components are being optimized to perform specific functions like storage\,\cite{Reagor_2013,Reagor_2016,Axline_2016,ganjam_surpassing_2024}, gates\,\cite{Niskanen_2007_coupler,Chen_2014,Moskalenko2022,Ding_2023_fluxonium,Marxer_2023transmon_coupler} or logical readout\,\cite{Jeffrey_2014,Walter_2017,Sunada_2024,Hazra_2024}. Depending on the desired functionality, different obstacles need to be overcome. These range from improving the noise resilience of qubits\,\cite{Pop2014,Hassani2023}, activating fast operations between circuit modes\,\cite{Xu_2020_coupler,Chapman_2023,Lu2023} and making them more resilient to strong driving\,\cite{Verney_2019,gusenkova_quantum_2021,Geisert_2024}. An interesting avenue to address these problems is to provide a linear inductive shunt for the nonlinear Josephson junction (JJ)\,\cite{Mooij_1999, Manucharyan_2009,Koch_2009,yan2020engineering,Ye_2021_Quarton,Kalacheva_2024}. This shunt confines the otherwise periodic Josephson potential while also enabling in-situ tunability of the circuit parameters with external magnetic fields\,\cite{Frattini_2017,Vool_2018,Lescanne2020_ATS}. Ideally, the inductor has a small footprint, making the circuit compact while simultaneously reducing the number of low frequency parasitic modes. Further, it has low internal loss, and is easy to fabricate and integrate with other circuit components (Fig.\,\ref{Fig1}a). While adding such an ideal linear inductor to the circuit QED toolkit would be very useful, it is challenging to realize.

Several approaches have already been utilized to approximate an ideal inductor, often in pursuit of making a superinductor\,\cite{Manucharyan_2009}. However, these approaches might not be optimal for building inductively shunted circuits with lower impedances. One approach leverages the geometric inductance of superconducting wires to build inductors that are highly linear\,\cite{Peruzzo_2020,Peruzzo_2021,Hassani2023}, but often requires the inductor to be several mm long in order to achieve a typical inductance of a few nH. This can introduce higher-order modes in the operational frequency range of the circuit which are often strongly coupled to the modes of interest. Another approach uses arrays of Josephson junctions (JJAs) to achieve a large inductance in a compact geometry\,\cite{Manucharyan_2009,Masluk_2012,Pop2014,Nguyen_2019,Kuzmin2019,Pechenezhskiy2020,Zhang_2021}. However, these inductors often have significant non-linearities and can also suffer from low frequency parasitic modes\,\cite{Ferguson_2013,Weissl_2015,Catelani_2015,DiPaolo2021}. In both cases, each additional parasitic mode increases the number of spurious interactions that can be accidentally activated during operation.

While both approaches used to realize a lumped-element, linear and low-loss inductor have limitations, a promising alternative approach is to utilize the high kinetic inductance in granular or disordered superconductors\,\cite{Annunziata_2010,ohya_room_2014,bruno_reducing_2015,dupre_tunable_2017,shearrow_atomic_2018,joshi_strong_2022,Frasca_NbN_2023,bahr_improving_2024,charpentier2024}. In particular, granular Aluminum (grAl) is an interesting candidate for this application as it can provide large sheet inductances of $\geq 2\,\mathrm{nH / sq}$\,\cite{grunhaupt_loss_2018,zhang_microresonators_2019,kristen_random_2023}, on par with lithographically defined JJAs. The microstructure of grAl consists of pure grains of aluminum (Al) that are separated by thin barriers of aluminum oxide\,\cite{Cohen_1968,Deutscher1973,Deutscher_1973_PII,yang_microscopic_2020} and can be modeled as an network of effective Josephson junctions (Fig.\,\ref{Fig1}b, \ref{Fig1}c) \,\cite{maleeva_circuit_2018}. Previous works have already shown that grAl can be used to make inductors with large inductances and low non-linearity\,\cite{maleeva_circuit_2018,grunhaupt_granular_2019,rieger_granular_2023}. However, single-photon quality factors of grAl resonators measured so far have been limited to $3\times10^5$ \cite{rotzinger_aluminium-oxide_2017, grunhaupt_loss_2018, maleeva_circuit_2018, zhang_microresonators_2019, borisov_superconducting_2020, he_lumped_2021}, precluding the use of grAl inductors in state-of-the-art circuits that require quality factors beyond $10^6$. 

In this work, we use $91 \pm 1 $ nm thick grAl films to build lumped-element linear inductors that are compatible with high coherence quantum circuits. Our smallest inductor, which is only 2 $\si{\micro \metre}$ wide and 30 $\si{\micro \metre}$ long, achieves an inductance of several nH and is fabricated using photolithography. To bound the non-linearity and microwave loss of grAl films of different sheet inductances, we characterize lumped-element resonators built by shunting the grAl inductors with in-plane capacitors. Our resonators have self-Kerr non-linearities of $0.2 - 20\,\mathrm{Hz/photon}$ indicating that the grAl inductors are much more linear than lithographically defined JJAs of similar size. Moreover, we demonstrate that our best resonators can have quality factors $> 3 \times × 10^6$, surpassing previously reported values for grAl resonators by an order of magnitude.

Further, we show that the grAl inductor can be integrated with other superconductors like Al or Ta to form the shunt capacitor, where the films are fabricated in separate steps and connected using an Al bandage layer. Our fabrication method alleviates design constraints that other in-situ fabrication techniques face and enables integration of grAl inductors in a wide variety of circuits. While we do not independently measure the losses associated with the contacts between grAl and the Al bandage, we find that they are sufficiently low loss such that the quality factor of our best hybrid grAl/Ta resonator surpasses $4.5\times 10^6$.

We also report on a previously unobserved trend that resonators built from grAl films with higher room-temperature resistivity systematically  have higher internal loss. After accounting for the dielectric loss in the substrate, we find that losses in resonators with the lowest resistivity films are comparable to surface losses in standard 3D Al transmons\,\cite{ganjam_surpassing_2024}. 
Additionally, our combined analysis of the all-grAl and hybrid samples suggests that the increased loss that we observe with film resistivity is in agreement with an increase in conductor losses in the grAl film. For the hybrid samples, the resistivity dependence could also have contributions from losses at the bandage contacts.  

Finally, to investigate the time stability of our grAl inductors, we monitor the resonance frequency of our devices. Occasionally, we find sudden shifts in the resonance frequencies that we attribute to the creation of excess quasiparticles following a high energy impact. Interestingly, we infer that the excess quasiparticle density relaxes back to its steady state value at time scales $\sim1.2\,\mathrm{ms}$ which are similar to values reported for Al transmons\,\cite{Wang2014} but three orders-of-magnitude faster than timescales previously reported for grAl films with similar resistivity\,\cite{grunhaupt_loss_2018,Henriques_2019}. Moreover, despite the more complicated superconducting gap profile in the hybrid resonators, we find similar relaxation times after an impact.

\section{Device design and measurement scheme}
Our strategy for quantifying the intrinsic loss in grAl inductors is to build lumped-element resonators entirely out of grAl and characterize them in a high-Q microwave environment. The resonator design is inspired by the 3D transmon geometry\,\cite{paik2011observation} which minimizes participation of the resonance mode in lossy dielectric interfaces ($p_{\mathrm{surf}} \sim 10^{-4}$) \cite{wang2015surface}. Further, the resonators are housed in a high-purity Al package that has four cylindrical tunnel waveguides. Since the resonator frequencies ($4 – 8\,\mathrm{GHz}$) are well below the waveguide cut-off frequency, they are insensitive to losses at the package seams\,\cite{Axline_2016, lei2023characterization, ganjam_surpassing_2024} (see App.\,\ref{appendix_package_loss}). Consequently, the total internal loss of the resonators is dominated by losses associated with the grAl thin films. 

\begin{figure}[t]
    \includegraphics[width =\linewidth]{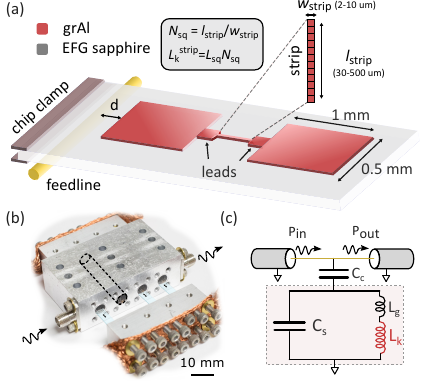}
    \caption{\textbf{All-grAl lumped-element resonator design and measurement scheme:} (a) Schematic illustrating the lumped-element resonator design used in this study. The design is inspired by the 3D transmon geometry (not to scale). The narrow strip of grAl is the dominant contributor to the total inductance of the resonator. (b) Photograph of a multiplexed coaxial tunnel package made from high-purity aluminum. The resonator chips (false-color coded in light blue) are clamped on one end and packaged into the cylindrical tunnels. (c) Effective circuit model of the hanger configuration used to characterize the resonator. $L_{\mathrm{k}}$ is the total kinetic inductance, which is dominated by the strip $L_{\mathrm{k}}^{\mathrm{strip}}$ but also contains contributions from the leads and the pads.
}
    \label{Fig_sample_design}
\end{figure}

In a typical 3D transmon, a single JJ provides a few nH of inductance and is shunted by a large capacitor to realize an operational frequency of a few GHz and an impedance of a few hundred ohm. Our resonators are a linear version of the transmon design with similar frequencies and impedances. Instead of a JJ, the inductance of these resonators primarily comes from a strip of grAl. As shown in Fig.\,\ref{Fig_sample_design}a, each resonator consists of a narrow strip of width $w_{\mathrm{strip}} = 2 - 10\,\si{\micro \metre}$ and length $l_{\mathrm{strip}} = 30 - 500\,\si{\micro \metre}$ made from grAl. The strip is connected via wide leads (width = $50\,\si{\micro \metre}$) to a pair of electrodes ($0.5\,$mm $\times$ $1\,$mm) forming the shunt capacitor. The length of the leads is adjusted to maintain a minimum separation of $300\,\si{\micro \metre}$ between the electrodes for each strip length. The total inductance of the resonator is dominated by the kinetic inductance $L_{\mathrm{k}}$ of the grAl film. The strip contributes a kinetic inductance $L^{\mathrm{strip}}_{\mathrm{k}} = L_{\mathrm{sq}}N_{\mathrm{sq}}$ where $L_{\mathrm{sq}}$ is the sheet inductance of the grAl film and $N_{\mathrm{sq}}  = l_{\mathrm{strip}}/w_{\mathrm{strip}}$ is the number of squares in the strip. The contribution of the strip to $L_{\mathrm{k}}$ is significantly larger than the pads or leads. Thus, the kinetic inductance can be varied by modifying $N_{\mathrm{sq}}$ or $L_{\mathrm{sq}}$ of the grAl film.  The frequency of the lumped-element resonator is determined by  ${\omega_r = 1/\sqrt{(L_{\mathrm{k}} + L_{\mathrm{g}})C_{\mathrm{s}}}}$ where $C_{\mathrm{s}}$ is the total capacitance and $L_{\mathrm{g}}$ is the stray geometric inductance.

For the all-grAl samples in this study, the entire resonator including the capacitor electrodes and leads are fabricated in a single photolithography step using direct laser writing (Heidelberg MLA 150). Each grAl film is deposited by zero-angle evaporation of pure Al (5N) in an oxygen environment, followed by liftoff. Adjusting the partial pressure of the oxygen in the chamber changes the normal-state resistivity $\rho_{\mathrm{n}}$ of the grAl film (see App.\,\ref{appendix_fab}). This in turn allows tuning of the films’s $L_{\mathrm{sq}}$  which is related to $\rho_{\mathrm{n}}$ by the Mattis-Bardeen theory\,\cite{mattis1958theory}.

The resonators are fabricated on unannealed 2$^{\prime\prime}$ EFG sapphire substrates. This substrate type has been previously characterized to have a bulk dielectric loss tangent of $(26.6\pm6.9) \times 10^{-8}$\,\cite{ganjam_surpassing_2024, read_precision_2023}. This results in bulk quality factors of $\sim 4.5 \times 10^6$ for our resonator geometries. For microwave characterization, the diced chips ($\sim 4 \times 20 \,\mathrm{mm}$) are clamped on one end using beryllium-copper leaf-springs and loaded into cylindrical tunnel waveguides made from superconducting high purity (5N5) aluminum as shown in Fig.\,\ref{Fig_sample_design}b. 

To characterize the resonance frequency $\omega_r$, self-Kerr $K$ and internal loss $Q_{\mathrm{int}}^{-1}$ of these lumped-element resonators, we capacitively couple the resonators to a common feedline and perform transmission measurements in a hanger configuration (Fig.\,\ref{Fig_sample_design}c). The coupling strength of each resonator to the feedline is characterized by an external quality factor $Q_{\mathrm{c}}$, which depends exponentially on the distance $d$\,($\sim 8 - 9\,$mm) between the resonator and feedline. For optimal loss characterization, the resonators are either critically coupled $Q_{\mathrm{c}} \sim Q_{\mathrm{int}}$ or under coupled $Q_{\mathrm{c}} > Q_{\mathrm{int}}$. This minimizes parasitic effects arising in the measurement lines which can lead to a distortion of the transmission coefficient\,\cite{Rieger_Fano_2023}. All resonators presented in this study are measured at the base temperature of {$T = 25\,\mathrm{mK}$} in a dilution refrigerator using a vector network analyzer (Agilent E5071C). The details of the measurement setup are summarized in App.\,\ref{appendix_setup}.

\begin{figure*}[t]
    \includegraphics[width =\linewidth]{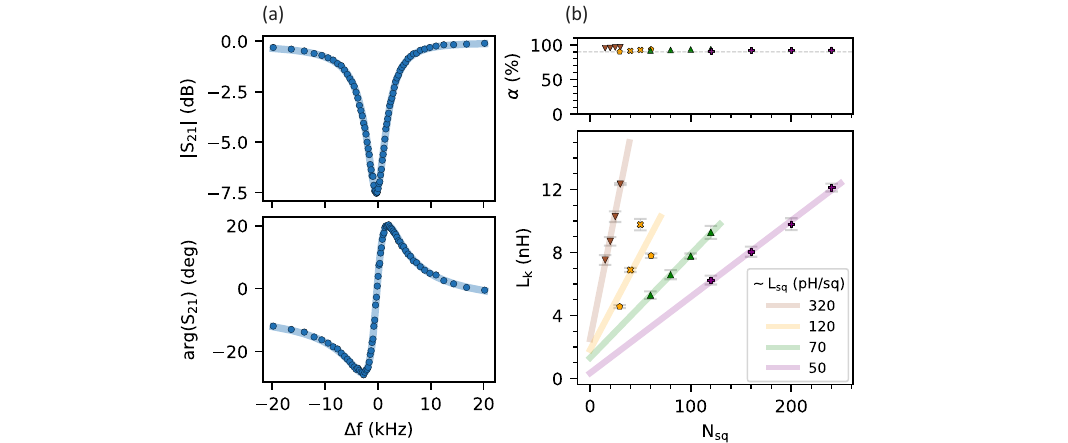}
    \caption{\textbf{Characterizing the kinetic inductance of the grAl films:} (a) Amplitude (top) and phase (bottom) of the transmission coefficient measured in hanger geometry at small drive powers ($\overline{n}_{\mathrm{ph}} \sim 1$) for an all-grAl resonator ($f_r$ = 6.04\,GHz, $\rho_{\mathrm{n}} \sim 2110\,\si{\micro \Omega}$cm, $l_{\mathrm{strip}} = 150\,\si{\micro \metre}$, $w_{\mathrm{strip}} = 3\,\si{\micro \metre}$) critically coupled to the feedline. Q$_\mathrm{i}$, Q$_\mathrm{c}$ and $\overline{n}_\mathrm{ph}$ are extracted using a circle fitting algorithm to fit the complex response. (b) Estimated kinetic inductance fraction $\alpha$ (top) and total kinetic inductance $L_{\mathrm{k}}$ (bottom) plotted as a function of $N_{\mathrm{sq}}$ in the grAl strip for a selection of samples (see Table\,\ref{table_info}). The different colors correspond to samples fabricated on the same wafer in a single round of fabrication. All resonators measured in this study have high kinetic inductance fractions ($>$ 90$\%$) while $L_{\mathrm{k}}$ can vary from 4 - 12 nH. The value of the sheet kinetic inductance $L_{\mathrm{sq}}$ is extracted from a linear fit and can be fine-tuned by controlling the oxygen content in the chamber during evaporation. 
}
    \label{Fig_Lkin}
\end{figure*}

\section{Characterizing kinetic inductance and self-Kerr non-linearity of all-grAl resonators}
\label{SEC_Lkin_Kerr}
To characterize our lumped-element resonators, we measure the frequency dependence of the complex transmission coefficient around the resonance frequency. The resonator response is given by\,\cite{Gao_2008, khalil2012analysis}: 
\begin{equation}
    S_{21}(\omega) = 1 - \frac{(Q_{\mathrm{L}}/|Q_{\mathrm{c}}|)e^{i\phi}}{1+i2Q_{\mathrm{L}}(\frac{\omega}{\omega_r}-1)}
    \label{eq_s21}
\end{equation}
where $\omega_r$ is the resonance frequency, $Q_{\mathrm{L}}^{-1} = Q_{\mathrm{int}}^{-1}+\Re[Q_{\mathrm{c}}^{-1}]$ is the loaded quality factor, $Q_{\mathrm{int}}$ is the internal quality factor and $Q_{\mathrm{c}} = |Q_{\mathrm{c}}|e^{-i\phi}$ is the complex external quality factor. The phase $\phi$ accounts for an asymmetry in the response caused by standing modes in the input lines\,\cite{khalil2012analysis}. $Q_{\mathrm{int}}$ and $Q_{\mathrm{c}}$ are extracted by fitting the linear response in the complex plane using a circle fit routine \cite{probst2015efficient}. For a hanger geometry, the average photon number in the resonator is $\overline{n}_{\mathrm{ph}} = (2Q_{\mathrm{L}}^2/\hbar \omega_r^2Q_{\mathrm{c}})P_{\mathrm{in}}$, where $P_{\mathrm{in}}$ is the input power at the chip level. The real and imaginary parts of the scattering coefficient $S_{21}$ measured for one such all-grAl resonator ($f_r$ = 6.04\,GHz, $\rho_{\mathrm{n}} \sim 2110\,\si{\micro \Omega}$cm, $l_{\mathrm{strip}} = 150\,\si{\micro \metre}$, $w_{\mathrm{strip}} = 3\,\si{\micro \metre}$), critically coupled to the feedline, are shown in Fig.\,\ref{Fig_Lkin}a. 
\subsection{Kinetic inductance}
From the measured resonance frequency, we can infer the kinetic inductance fraction $\alpha =1 -L_{\mathrm{g}} C_{\mathrm{s}}\omega_r^2$ and the total kinetic inductance $L_{\mathrm{k}}= \alpha L_{\mathrm{g}}/(1-\alpha)$ for each device. Here $L_{\mathrm{g}}$ and $C_{\mathrm{s}}$ are estimated by simulating the resonator using an FEM  electromagnetics solver (Ansys HFSS). For our resonator geometries, $C_{\mathrm{s}}$ is between $\sim 84 - 100\,\mathrm{fF}$ and $L_{\mathrm{g}}$ can vary from $\sim 360 - 930\,\mathrm{pH}$ with strip dimension. In combination with the kinetic inductance, this results in a circuit impedance of $200-400\,\Omega$. Fig.\,\ref{Fig_Lkin}b summarizes the $\alpha$ (top) and $L_{\mathrm{k}}$ (bottom) measured for resonators fabricated with different $N_{\mathrm{sq}}$ in the strip. The different colors correspond to samples fabricated on the same wafer in the same round of lithography. For each wafer, a $91\pm 1\,\mathrm{nm}$ thick grAl film was deposited by evaporating Al at $1\,\mathrm{nm/s}$ under a different oxygen partial pressure ($p_{\mathrm{ox}}$) to vary $L_{\mathrm{sq}}$ (see Fig.\,\ref{fig_rho_vs_O2}). The typical value of $p_{\mathrm{ox}}$ in these depositions was $\sim 5\times 10^{-5}\,\mathrm{mbar}$ and was obtained dynamically by controlling the oxygen flow rate while simultaneously pumping on the system. To improve uniformity of the film resistivity across the wafer, the substrate was rotated at a constant speed of 25$\,$deg/sec during deposition. As expected, we find the $\alpha$ for all our resonators is close to unity. Further, to achieve resonator frequencies in the $4-8\,\mathrm{GHz}$ range, we realize $L_{\mathrm{k}}$ of $4 - 12\,\mathrm{nH}$. Using a linear fit to the data we extract the sheet inductance of each film, which varies from $\sim 30\,\mathrm{pH/sq}$ to $\sim 320\,\mathrm{pH/sq}$ between the different wafers. Thus, despite using relatively thick grAl films, we can still realize very compact inductors. This contrasts with some other kinetic inductance superconductors where the films must be made very thin to achieve comparable sheet inductances. Although we do not explore higher sheet inductances in this study, $L_{\mathrm{sq}}$ values as high as 2 nH/sq can be achieved in grAl films by further increasing the oxygen content or by reducing the film thickness\,\cite{grunhaupt_loss_2018}.
\subsection{Self-Kerr non-linearity and critical current density}
\begin{figure}[t]
    \includegraphics[width =\linewidth]{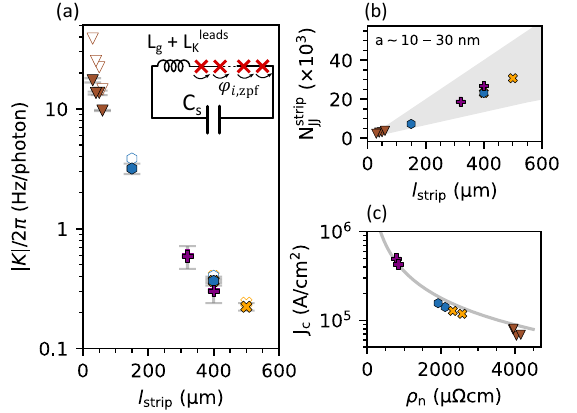}
    \caption{\textbf{Self-Kerr coefficients of the all-grAl resonators:} (a) Magnitude of the self-Kerr coefficient $K$ (solid markers) of each resonator  plotted vs grAl strip length $l_{\mathrm{strip}}$. The self-Kerr is extracted from the power dependence of the resonance frequency. From the energy participation ratio of the strip, we infer the self-Kerr coefficient associated with just the strip (open markers). The difference between the measured and inferred values is negligible for the longest strips in our geometry. The strip can be modeled as a 1D array of effective JJs with an equal phase drop $\varphi_{i,\mathrm{zpf}}$ across each junction. (inset) (b) Number of effective JJs in the strip ($N^{\mathrm{strip}}_{\mathrm{JJ}}$) calculated using the inferred self-Kerr coefficient and the charging energy of the shunt capacitance $C_{\mathrm{s}}$. These values are consistent with an effective JJ size of $\sim 10 - 30\,$nm (shaded region). (c) Using this model, we estimate the inductance associated with each effective JJ and infer a critical current of the strip. The resulting critical current density $J_{\mathrm{c}}$ is plotted as a function of normal state resistivity of the grAl film. 
    The solid line is a guide to the eye for a $1/\rho_{\mathrm{n}}$ dependence. 
}
    \label{Fig_Kerr}
\end{figure}

To probe the effects of the intrinsic non-linearity associated with the kinetic inductance of grAl, we quantify the self-Kerr coefficient $K$ of our resonators. To extract $K$, we measure the power dependent shift of the resonator frequency.  The magnitude of $K$ for the all-grAl resonators is plotted in Fig.\,\ref{Fig_Kerr}a. We find that the resonators are all weakly non-linear with $K$ varying from $0.2-20\,\mathrm{Hz/photon}$, substantially lower than the linewidths. The photon number in the resonator was calculated based on a room temperature calibration of the input power and all samples were measured using the same line configuration as shown in App.\,\ref{appendix_setup}.  While the total kinetic inductance depends on both $L_{\mathrm{sq}}$ and $N_{\mathrm{sq}}$, we find that the self-Kerr coefficient for these devices simply scales with the length of the strip. 

By modelling the grAl inductor strip as a one-dimensional array of effective JJs \cite{maleeva_circuit_2018}, we relate $K$ with the geometry of the resonator. Assuming that the total phase drop along the inductor is distributed homogeneously across each effective junction in the array, the self-Kerr coefficient can be approximated as $E_{\mathrm{c}}/N_{\mathrm{JJ}}^2$\,\cite{Eichler2014,winkel_implementation_2020}. Here $E_{\mathrm{c}} = e^2/(2C_\mathrm{s})$  is the charging energy of a single electron associated with the shunt capacitance of the resonator and $N_{\mathrm{JJ}}$  is the total number of effective junctions in the array. As compared to a transmon qubit with the same capacitor geometry, for which the anharmonicity is approximately given by $E_{\mathrm{c}}$\,\cite{Koch_2007}, the non-linearity of our lumped-element resonator is effectively diluted by a factor of $N_{\mathrm{JJ}}^2$. Our measured small non-linearities thus imply that the $N_{\mathrm{JJ}}$ in our inductor strips is very large. Further, we can relate $N_{\mathrm{JJ}}$ to our resonator geometry as $N_{\mathrm{JJ}}=l_{\mathrm{strip}}/a$ where $a$ is the effective size of a single junction in this model (see Fig.\,\ref{Fig1}c). This results in the relation $|K| = p^2E_{\mathrm{c}} a^2/l_{\mathrm{strip}}^2$ between the measured self-Kerr coefficient and the model parameters. Here $p$ is the energy participation ratio of the strip and accounts for the reduction of the measured non-linearity due to additional stray inductance from the leads\,\cite{Minev2021}. This modification (difference between filled and open markers in Fig.\,\ref{Fig_Kerr}a) is especially important for shorter strip lengths in order to accurately estimate $N_{\mathrm{JJ}}$ (see App.\,\ref{App_Kerr_correction}). 

Using this simple phenomenological model, we can estimate $N_{\mathrm{JJ}}$ in our inductor strips as shown in Fig.\,\ref{Fig_Kerr}b from the measured self-Kerr coefficients. For all devices, we individually simulate the charging energy $E_{\mathrm{c}}$ associated with the large shunt capacitance $C_\mathrm{s} \approx 84 - 100\,\mathrm{fF}$ and estimate the stray inductance arising in the pads and the leads from the geometry. We find $N_{\mathrm{JJ}} \sim$ 4000 - 30,000 in our grAl resonators with strip lengths of only a few 100 um, making them substantially more linear than JJA resonators of similar length. Further, the estimated $N_{\mathrm{JJ}}$ increases approximately linearly with the strip length, consistent with an effective junction size of $a \sim 10 - 30\,\mathrm{nm}$\,\cite{yang_microscopic_2020}, as indicated by the grey cone in Fig.\,\ref{Fig_Kerr}b. 

From the total number of junctions and the kinetic inductance associated with the strip, we also estimate the inductance and critical current of the effective JJs in our model as $L_{\mathrm{J}} = L_{\mathrm{K}}^{\mathrm{strip}}/N_{\mathrm{JJ}}$ and $I_{\mathrm{c}} = \Phi_0/(2\pi L_{\mathrm{J}})$, respectively. Further, we use this critical current along with the strip cross section $A = w t$ to infer a critical current density $J_{\mathrm{c}} = I_{\mathrm{c}} / A$. Since the thickness of our grAl films $t$ and the width of the strips $w_{\mathrm{strip}}$ are smaller than the penetration depth $\lambda \gtrsim 1 \,\si{\micro\metre}$\,\cite{Cohen_1968} and the perpendicular penetration length $\lambda_{\perp} (\sim \lambda^2/t > 12.5\, \si{\micro \metre}$), respectively, we can assume a uniform current density along the strip's cross section. The resulting critical current density is shown in Fig.\,\ref{Fig_Kerr}c as a function of film resistivity. We find that $J_\mathrm{c} \propto \rho_{\mathrm{n}}^{-1}$ scales inversely with the normal-state resistivity and is $\sim 10^5\,-10^6$\,A/cm$^2$ consistent with transport measurements on thinner grAl films with similar resistivity\,\cite{Friedrich_2019,Winkel_PhDthesis}. Interestingly, these values are only one to two orders of magnitude lower than critical current densities measured for pure Al ($J_{\mathrm{c}} > 10^7$\,A/cm$^2$ \cite{romijn1982critical}) but three to four orders of magnitude higher than conventional Al/$\mathrm{AlO}_x$/Al JJs ($J_{\mathrm{c}} \sim 100$\,A/cm$^2$). This observation poses questions about the type of these effective junctions.

Notably, it has been shown that JJs made from small volumes of grAl with dimensions comparable to the effective junction size are likely to have a sinusoidal current-phase relation\,\cite{rieger_granular_2023}. However, due to the large number of effective junctions in our resonators we are only weakly probing the non-linearity of the intergrain coupling in grAl. Therefore, we cannot distinguish whether the microstructure of grAl consists of grains fully surrounded by insulating AlO$_{\mathrm{x}}$ coupled by the Josephson effect or is a network of weak links. Yet, we find that this model describes the non-linear properties of our resonators quite well. 

\section{All-grAl resonators}
\label{sec_loss_all_grAl}
In the previous sections, we have shown that our grAl inductors can be made compact and linear.  To validate their use in circuits, we also need to demonstrate that these inductors allow high circuit coherence. Conductor loss in the grAl film and dielectric loss at the grAl interfaces could lower the device quality factor compared to state-of-the-art materials\,\cite{McRae_2020}. To investigate these losses, we extract the internal quality factor of the all-grAl resonators from the measured transmission coefficient (Eq. \ref{eq_s21}). 

\begin{figure}[t]
    \includegraphics[width =\linewidth]{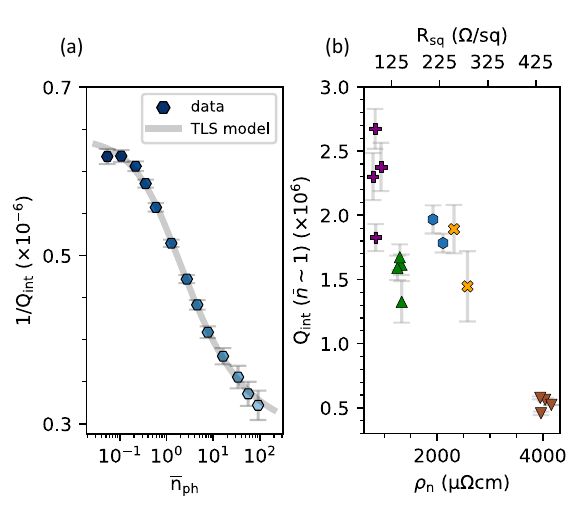}
    \caption{\textbf{Measured internal loss in the all-grAl resonators:}
     (a) Power dependence of the measured internal loss $1/Q_{\mathrm{int}}$. At single photon powers, this resonator ($f_r$ = 6.04\,GHz, $\rho_{\mathrm{n}} \sim 2110\,\si{\micro \Omega}$cm, $l_{\mathrm{strip}} = 150\,\si{\micro \metre}$, $w_{\mathrm{strip}} = 3\,\si{\micro \metre}$) has a $Q_\mathrm{int}$ of 1.86 $\pm$ 0.01 x 10$^6$. Beyond $\overline{n}_\mathrm{ph} \sim$ 100 the resonator approaches bifurcation. $1/Q_\mathrm{int}$ monotonically decreases with power and the fit corresponds to a saturable two-level system model that results in power-dependent loss. (b) Internal quality factors for various all-grAl resonators measured in this study plotted as a function of the measured room temperature resistivity of the all-grAl film. The corresponding sheet resistance is indicated on the top axis. The grAl film is $91\pm1$ nm thick for all resonators. The error bars are estimated from the standard deviation of $Q$s obtained from repeated measurements (see App.\,\ref{appendix_Q_fluct}).}
    \label{Fig_Qs}
\end{figure}

The extracted total internal loss $1/Q_{\mathrm{int}}$ is shown in Fig.\,\ref{Fig_Qs}a as a function of input drive power for a critically coupled resonator. We find that the measured loss is power dependent and decreases monotonically with $\overline{n}_{\mathrm{ph}}$. At higher powers ($\overline{n}_{\mathrm{ph}} > 100$), the response becomes highly non-linear, and we do not approach a high-power $Q_{\mathrm{int}}$ limit before bifurcation. The observed power dependence indicates that the loss is from a saturable environment. While this behavior is well captured by a generic two-level system model (solid line in Fig.\,\ref{Fig_Qs}a, see App.\,\ref{TLS_model}), our measurement cannot identify the origin of this environment. Therefore, we use the fit to the power dependence only as an interpolating function to extract the internal quality factor in the single photon regime. For this resonator, we find the $Q_{\mathrm{int}}$ to be $\sim 1.8\times 10^6$ which, significantly exceeds the highest single-photon $Q_{\mathrm{int}}$ reported for any resonator made out of a grAl thin film \cite{rotzinger_aluminium-oxide_2017, grunhaupt_loss_2018, maleeva_circuit_2018, zhang_microresonators_2019, borisov_superconducting_2020, he_lumped_2021}. Notably, this is not the highest internal quality factor we have measured in this study.

To investigate if the normal-state resistivity of the grAl film affects the internal loss, we measure the single-photon $Q_{\mathrm{int}}$ of several all-grAl resonators fabricated under different oxygen partial pressures. Our highest resistivity films remain well below the superconductor-insulator transition limit of grAl ($\rho_{\mathrm{n}}>10^5$ $\si{\micro \Omega\centi \metre}$\,\cite{Dynes_1984, Bachar_2015}). In Fig.\,\ref{Fig_Qs}b we plot $Q_{\mathrm{int}}$ versus the measured normal-state resistivity $\rho_{\mathrm{n}}$. Resonators plotted in the same color are samples from the same wafer and should ideally have the same resistivity. However, we find that the resistivity of the film can have a radial gradient of $\sim 7\,\%$ across the wafer. Therefore, for each resonator, $\rho_{\mathrm{n}}$ is inferred from four-probe van der Pauw measurements of test structures patterned in its close vicinity. 

Interestingly, we find a strong inverse correlation of the $Q_{\mathrm{int}}$ with $\rho_{\mathrm{n}}$ and $R_{\mathrm{sq}} (=\rho_{\mathrm{n}}/t$). As shown in Fig.\,\ref{Fig_Qs}b, $Q_{\mathrm{int}}$ degrades by a factor of $\sim 6$ (from a maximum of $2.67 \times 10^6$ to a minimum of $0.46\times 10^6$) as $\rho_{\mathrm{n}}$ increases from $\sim\,830\,\si{\micro \Omega}$cm to $\sim\,3970\,\si{\micro \Omega}$cm. Since all the resonators have the same film thickness, our finding implies that $Q_{\mathrm{int}}$ degrades with increasing $L_{\mathrm{sq}}$. 
Therefore, there is a price to pay in quality as we attempt to make the grAl inductor more compact. In future studies, it would be interesting to investigate the relationship between film thickness $t$ and $Q_{\mathrm{int}}$ for grAl devices with identical geometry, to understand whether the increase in loss is related to the sheet resistance or the film resistivity.

\begin{figure*}[t]
    \includegraphics[width =0.975\linewidth]{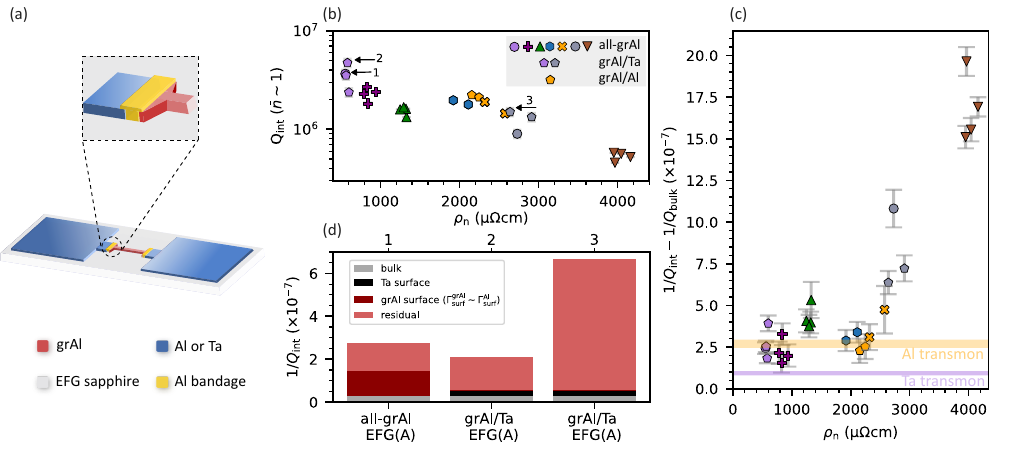}
    \caption{\textbf{Hybrid lumped-element resonators:} (a) Schematic illustration of the hybrid resonator design. The inductor strip (red) is made out of grAl while the connecting leads and capacitor pads (blue) are made out of pure Al or Ta. The materials are integrated ex-situ using a bandage layer (yellow) made out of pure Al. (b) Internal quality factors of the hybrid and all-grAl resonators measured at single photon powers. The hybrid resonators have quality factors comparable to the all-grAl resonators fabricated on the same wafers and follow a similar trend with resistivity. (c) The residual loss ($1/Q_{\mathrm{int}} - 1/Q_{\mathrm{bulk}}$) of all resonators plotted as a function of grAl film resistivity. The horizontal orange and violet lines represent the residual loss estimated for a typical 3D Al transmon and Ta transmon, respectively. The residual losses for the low-resistivity all-grAl resonators are comparable to that of the Al transmons with similar $p_{\mathrm{surf}}$, indicating that $\Gamma_{\mathrm{surf}}^{\mathrm{grAl}}\sim \Gamma_{\mathrm{surf}}^{\mathrm{Al}}$ at these resistivities. The hybrid grAl/Ta resonators do not reach the Ta transmon limit indicating that they are not limited by surface losses. (d) Loss budget for the four samples indicated by the arrows in panel (b) distinguishing between bulk dielectric loss, surface loss and residual losses. The surface loss due to grAl is calculated assuming a surface loss factor of pure Al. For the hybrid grAl/Ta samples, the contribution of the grAl film to the total surface loss is negligible.}
    \label{Fig_hybrid}
\end{figure*}

\section{Hybrid resonators}
\label{SEC_hybrid_resonators}
With our all-grAl resonators, we have shown that grAl inductors can be made compact, linear and low-loss. However, to use grAl inductors in state-of-the-art circuits, they must be integrable with conventional Al/AlO$_{\mathrm{x}}$/Al JJ fabrication, or with materials that have lower surface dielectric loss factors like Ta\,\cite{place2021new,ganjam_surpassing_2024,Wang2022,Crowley_2023}, Nb\,\cite{Alto_2022,Bal2024} and Re\,\cite{Wang_Re_2009,Sage_Re_2011}. Prior efforts to implement circuits using grAl and Al have leveraged the fact that both superconductors have compatible fabrication processes. This enables in-situ contact between the two materials using angled depositions. However, this method imposes practical constraints on the circuit geometries. Typically, it uses electron-beam lithography and requires the inductor to be narrow on the order of a few hundred nanometers\,\cite{grunhaupt_granular_2019,Geisert_2024}. Here we alleviate these constraints by using an ex-situ contact between grAl and Al while using only zero-angle deposition in combination with photolithography.

To test this contact, we build hybrid lumped-element resonators that have the same geometry as the all-grAl samples described earlier, but with capacitor electrodes made from Al (Fig.\,\ref{Fig_hybrid}a). We use a multi-step lithography process to fabricate these hybrid resonators. In the first step, we only fabricate the inductor strip out of $91\pm 1 \, \mathrm{nm}$ thick grAl. In the second step we fabricate the Al capacitor electrodes, $80\,\mathrm{nm}$ thick, without making contact to the grAl strip. In a third step, we lithographically define a small contact region and use Ar ion milling\,\cite{grunhaupt2017argon} to remove the native oxides on the surfaces of both films. Finally, we deposit a $150\,$nm thick Al bandage in this region to make contact. Although we could directly make contact between grAl and Al in the second step, using a bandage layer has advantages. It not only minimizes damage to the substrate from ion milling \cite{dunsworth2017characterization}, it also enables contact to double-angle deposited Josephson junctions and other superconductors, even if they are sputtered or epitaxially grown.

To illustrate this advantage of using a dedicated bandage layer, we also build hybrid resonators where the capacitor electrodes are made out of Ta (App.\,\ref{appendix_fab}). In order to benefit from its low surface dielectric loss factor, these devices are fabricated on annealed EFG substrates. The bulk dielectric loss factor for this substrate type is ($3.64\pm2.5)\times 10^{-8}$ \cite{ganjam_surpassing_2024}, resulting in an average bulk quality factor of ${\sim 33 \times 10^6}$ for our resonator geometries. In Fig.\,\ref{Fig_hybrid}b, we show the single-photon quality factors of the grAl/Al and grAl/Ta hybrid resonators fabricated using this bandage technique together with the all-grAl devices. The highest quality factor we measure is $4.7\times10^6$ for a hybrid grAl/Ta sample at $\rho_n \sim 580\,\si{\micro\Omega\centi\metre}$. Interestingly, we find that the hybrid resonators have quality factors comparable to the all-grAl resonators fabricated on the same wafers and follow a similar trend with resistivity. This is in contrast with the improvement observed in transmon qubits when replacing the Al electrodes with Ta \cite{ganjam_surpassing_2024}, suggesting that our hybrid samples may not be limited by surface losses. 

\section{Loss analysis}
To further quantify the losses solely associated with the grAl films at each resistivity, we account for the role played by the geometry of the devices. We do so by using a generalized energy participation ratio model similar to Ref. \cite{wang2015surface,ganjam_surpassing_2024}. Since our package loss is negligible (see App.\,\ref{appendix_package_loss}), we attribute the total internal loss to four distinct loss channels:

\begin{equation}
    1/Q_{\mathrm{int}} =  1/Q_{\mathrm{bulk}} +  1/Q_{\mathrm{surf}} + 1/Q_{\mathrm{ind}} + 1/Q_{\mathrm{contact}}
\end{equation}

Here the first term $Q_{\mathrm{bulk}}^{-1} = p_{\mathrm{bulk}}\Gamma_{\mathrm{bulk}}$ represents the loss in the substrate dielectric, characterized by a bulk dielectric loss factor $\Gamma_{\mathrm{bulk}}$ and an energy participation ratio $p_{\mathrm{bulk}}$. For unannealed EFG sapphire, $\Gamma_{\mathrm{bulk}}$ has been measured independently by \cite{read_precision_2023, ganjam_surpassing_2024} to be $(26.6\pm6.9) \times 10^{-8}$ on average. The second term $Q_{\mathrm{surf}}^{-1} =p_{\mathrm{surf}}\Gamma_{\mathrm{surf}}$ is associated with loss in the dielectrics ($t_{\mathrm{diel}} = 3$\,nm, $\epsilon_r = 10$) at the metal-air, metal-substrate and substrate-air interfaces in the device. The corresponding energy participation ratio is $p_{\mathrm{surf}} = p_{\mathrm{MA}} + p_{\mathrm{MS}} + p_{\mathrm{SA}}$ and $\Gamma_{\mathrm{surf}}$ is the surface loss factor. The geometric participation ratios are estimated by using a two-step FEM simulation \cite{wang2015surface}. For our devices, the dominant contribution to $p_{\mathrm{surf}}$ is from the capacitor electrodes and leads ($>94\,\%$) while the contribution of the grAl strip is significantly smaller ($< 6\,\%$). The third term $Q_{\mathrm{ind}}^{-1}$ accounts for additional conductor loss in the grAl strip that may be caused by an excess background of quasiparticles (QPs) \,\cite{Zmuidzinas_2012} or by any other loss mechanism that also couples inductively. The fourth term $Q_{\mathrm{contact}}^{-1}$ accounts for conductor loss associated with the contacts formed by the bandage in the hybrid samples. 

Since our samples are made on two different sapphire substrates, we subtract the contribution of the bulk dielectric loss from the measured internal loss. We refer to the remaining losses as ``residual loss'' $Q_{\mathrm{res}}^{-1} = Q_{\mathrm{int}}^{-1} - Q_{\mathrm{bulk}}^{-1}$. The estimated residual losses for all samples are plotted in Fig.\,\ref{Fig_hybrid}c. We see that the residual loss is significantly higher for the high resistivity samples. For reference, we also plot the residual loss estimated for an Al transmon (orange line) and a Ta transmon (violet line)\,\cite{ganjam_surpassing_2024} which are known to be limited by surface losses. Interestingly, we find that the residual losses for the low-resistivity all-grAl resonators are similar to that of the Al transmons. Since the $p_{\mathrm{surf}}$ of grAl $(0.6- 0.9) \times 10^{-4}$ in these resonators is similar to the $p_{\mathrm{surf}}$ of Al in a typical Al transmon ($\sim 1.5 \times 10^{-4})$\,\cite{wang2015surface, ganjam_surpassing_2024}, we conclude that the surface loss factor $\Gamma_{\mathrm{surf}}^{\mathrm{grAl}}$ of low-resistivity grAl films is similar to pure Al. Additionally, a previous loss characterization of all-grAl resonators has estimated a surface loss factor comparable to Al even with $\rho_{\mathrm{n}} \sim 2000-3000\,\si{\micro\Omega}$cm \cite{grunhaupt_loss_2018}. This would imply that $\Gamma_{\mathrm{surf}}^{\mathrm{grAl}}$ doesn't change significantly in the resistivity range we investigate. Since the all-grAl resonators don't have any additional contacts ($Q_{\mathrm{contact}}^{-1}=0$), we attribute their increasing losses with resistivity to an increase in conductor loss in the grAl inductor strip.

For the hybrid grAl/Ta samples, since the surface loss factor associated with the Ta electrodes and leads is significantly smaller than Al and grAl, we would expect these resonators to have residual losses comparable to typical 3D Ta transmons (violet line in Fig.\,\ref{Fig_hybrid}c).
However we find that their residual losses are actually comparable to the all-grAl resonators of similar resistivities. This suggests that $Q_{\mathrm{ind}}^{-1} + Q_{\mathrm{contact}}^{-1}$ dominates over $Q_{\mathrm{surf}}^{-1}$ in these devices, as illustrated in the loss budget shown in Fig.\,\ref{Fig_hybrid}d. This conclusion is in agreement with increasing conductor loss at higher resistivities. However, we cannot exclude the possibility that the losses at the contacts may also have a resistivity dependence. 

In order to quantitatively disentangle the individual contributions of these loss channels, a more detailed multimodal analysis would be required \cite{lei2023characterization, ganjam_surpassing_2024} that also includes conductor loss. Microscopically, what physical mechanism gives rise to the increase in conductor loss remains an open question. In grAl, deviations from the BCS model have been observed in the form of sub-gap states\,\cite{yang_microscopic_2020} and increased broadening of the superconducting gap with normal-state resistivity\,\cite{Dynes_1984}. Independent of the physical origin, our finding suggests the use of low-resistivity grAl films in circuits that are more sensitive to conductor loss, for instance in flux-qubits. On the other hand, high resistivity grAl films are suitable for applications where achieving a high sheet inductance is more important than the circuit coherence, for example in the readout of spin qubits\,\cite{Stockklauser_2017,Landig_2018,Samkharadze_2018, Scarlino_2019,Boettcher2022,Oakes_2023_Spin, janík2024_grAl_spinqubit}. 

\begin{figure*}[t]
    \includegraphics[width = 0.85\linewidth]{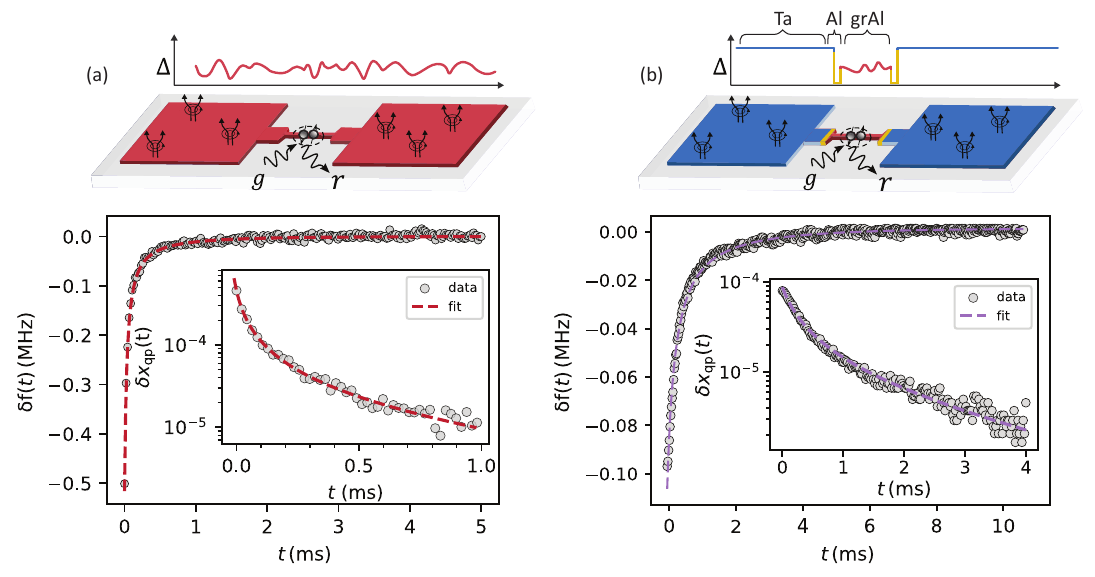}
    \caption{\textbf{QP time dynamics after high energy impact:} (a) Frequency response of an all-grAl resonator ($f_{\mathrm{r}} =$ 4.60\,GHz, $l_{\mathrm{strip}} =60\,\si{\micro\metre}$, $w_{\mathrm{strip}} =2\, \si{\micro\metre}$, $\rho_{\mathrm{n}}\sim 4160\,\si{\micro\Omega}$cm, $Q_{\mathrm{c}}\sim 400$) following a high-energy impact event measured at $\overline{n}_{\mathrm{ph}} \sim 10^5$. The frequency drop  is followed by an initial fast recovery and then an exponential tail as it re-approaches steady state. The corresponding change in quasiparticle density $\delta x_{\mathrm{qp}}$ following the impact is shown in the inset. The time dynamics are well described by a model that accounts for both recombination and trapping (red dashed line) with an exponential decay constant $\tau_\mathrm{ss} = 1.2 \pm 0.1\,$ms, a recombination constant $r = 1/(16 \pm 3\,\mathrm{ns})$, a trapping rate $s=1/(1.3 \pm 0.2\,\mathrm{ms})$ and a generation rate $g = (6.5\pm0.7) \times10^{-4}$\,s$^{-1}$. The trapping could be due to vortices in the pads or spatial variations in the superconducting gap ($\Delta$) of grAl as illustrated in the top panel. (b) Frequency response and $\delta x_{\mathrm{qp}}$ of a grAl/Ta resonator with Al bandage contacts ($f_{\mathrm{r}} =$ 5.19\,GHz, $l_{\mathrm{strip}} =200\,\si{\micro\metre}$, $w_{\mathrm{strip}} =5\, \si{\micro\metre}$, $\rho_{\mathrm{n}}\sim 3090\,\si{\micro\Omega}$cm, $Q_{\mathrm{c}}\sim 10^4$) following a high-energy impact event measured at $\overline{n}_{\mathrm{ph}} \sim 10^4$. We extract an exponential decay constant $\tau_\mathrm{ss} = (2.8 \pm 0.1)\,  \mathrm{ms} $, recombination rate $r \sim 1/(30 \pm 3\,\mathrm{ns})$, trapping rate $s = 1/(2.9 \pm 0.3\,\mathrm{ms})$, and generation rate $g = (2.0\pm0.5) \times10^{-4}\, $s$^{-1}$. Interestingly, despite the significantly larger gap of the Ta electrodes, these values are similar to those extracted for the all-grAl device. In both cases, the relaxation times are orders of magnitude faster than previous reports on grAl resonators. }
    \label{Fig_impacts}
\end{figure*}

\section{Quasiparticle dynamics}
\label{subsec_QP}

In the previous sections, we have shown that our lumped-element grAl inductors are simultaneously linear, low loss and readily integrable with other superconductors. However, their use in state-of-the-art quantum circuits also requires the resonance frequency to remain stable with time. In general, excess QPs in the inductor caused by background radiation or cosmic ray impacts\,\cite{Day2003,Karatsu_2019,Vepsalainen2020,Cardani2021,Gusenkova_2022_underground,harrington2024} could result in frequency shifts. In particular, previous measurements on grAl resonators have observed seconds long QP relaxation times after high energy impacts\cite{grunhaupt_loss_2018}. This time scale is significantly longer than relaxation times observed in other superconductors \cite{barends2008QPrelax} and would limit the applicability of grAl in quantum information processing schemes where frequency stability is important.

To investigate QP lifetimes in our grAl resonators, we monitor the resonator response to a high energy impact. Fig.\,\ref{Fig_impacts}a shows the response of an all-grAl resonator measured at $\sim 10^5$ photons, ($f_{\mathrm{r}} =$ 4.60\,GHz, $l_{\mathrm{strip}} =60\,\si{\micro\metre}$, $w_{\mathrm{strip}} =2\, \si{\micro\metre}$, $\rho_{\mathrm{n}}\sim 4160\,\si{\micro\Omega}$cm, $Q_{\mathrm{c}}\sim 400$) immediately following such an event. We see a sudden drop in frequency of the resonator due to the rapid increase of excess QPs generated by the impact. After an initial fast recovery within the first tens of microseconds, the resonator frequency relaxes back to steady state exponentially in a few milliseconds. Interestingly, this relaxation time is significantly ($\sim 1000 \times$) shorter than the values previously reported in stripline grAl resonators with similar resistivity made with much thinner films \cite{grunhaupt_loss_2018}. 

To extract the non-equilibrium QP relaxation time $\tau_{\mathrm{ss}}$, we use a phenomenological model similar to Ref.\,\cite{Wang2014} that considers three distinct processes: recombination with other QPs with recombination constant $r$, trapping at rate $s$ and a constant background generation rate $g$ (see App.\,\ref{appendix_impacts_procedure}). 
\begin{equation}
    \frac{dx_{\mathrm{qp}}}{dt} = -rx^2_{\mathrm{qp}} -sx_{\mathrm{qp}} + g
\end{equation}

We find that our time constant for the exponential tail $\tau_{\mathrm{ss}}= 1.2 \pm 0.1\,$ms  is very similar to the value reported for standard 3D Al transmons that were dominated by trapping, potentially due to vortices in the capacitor pads\,\cite{Wang2014}.  For the impact shown in Fig.\,\ref{Fig_impacts}, we obtain $r =1/(16 \pm 3\,\mathrm{ns})$, $s =1/(1.3 \pm 0.2\,\mathrm{ms})$ and $g=(6.5\pm 0.7)\times 10^{-4}$\,s$^{-1}$ assuming a steady state $x^0_{\mathrm{qp}} = 8 \times 10^{-7}$ estimated from the measured internal loss. Given that we expect the diffusion constant of grAl to be significantly smaller than in pure Al due to the higher resistivity, the trapping rate we find in our samples might be determined by a different mechanism. For example, local variation of the gap could also cause QP trapping in addition to vortices in the pads\,\cite{deGraaf_2020_qTLS}. 

In the hybrid grAl/Ta resonators, since the superconducting gap of Ta is much higher than Al or grAl it could negatively affect the relaxation of QPs in these devices. To investigate this, we also study QP dynamics in a hybrid grAl/Ta sample ($f_{\mathrm{r}} =$ 5.19\,GHz, $l_{\mathrm{strip}} =200\,\si{\micro\metre}$, $w_{\mathrm{strip}} =5\, \si{\micro\metre}$, $\rho_{\mathrm{n}}\sim 3094\,\si{\micro\Omega}$cm, $Q_{\mathrm{c}}\sim 10^4$) at $\sim 10^4$ photons. The observed frequency shift and change in $x_{\mathrm{qp}}$ is shown in Fig.\,\ref{Fig_impacts}b. Interestingly, we observe similar time constants for the relaxation of QPs in this hybrid device as in the all-grAl resonator. While the trapping mechanisms at play need further investigation, this observation proves that interfacing grAl with a high-gap superconductor using an Al bandage need not impact the QP dynamics negatively. Consequently, our results emphasize that long QP lifetimes are not the inevitable fate of grAl inductors.

\section{Discussion/Conclusions}
In summary, using the high kinetic inductance of grAl, we have realized an inductor that is simultaneously lumped-element, linear and low-loss. We have fabricated $\sim 90\,\mathrm{nm}$ thick grAl films with sheet inductances of $30 - 320$ pH/sq to implement grAl inductors with several nH of inductance. To validate their utility in high-coherence circuits, we have characterized a total of 20 all-grAl resonators. Their low self-Kerr non-linearity ranging between 0.2 and 20\,Hz/photon makes them significantly more linear than conventional arrays of Josephson junctions of similar length. We find that our data is well described by an effective circuit model that treats the grAl inductor as a 1D array of effective JJs. We estimate the size of these effective junctions to be around 10 - 30\,nm, corresponding to a large number of JJs even for micrometer sized elements. 

Further, by taking advantage of the 3D transmon geometry and the low loss environment provided by our coaxial tunnel package, we demonstrate internal quality factors that are well beyond $10^{6}$. Since our measurements are primarily sensitive to losses related to the grAl film and the substrate, we uncover a correlation between the normal-state resistivity of grAl and microwave losses. This implies a trade-off between quality factor and resistivity of the grAl film, suggesting that resistivity is an important design consideration for grAl lumped-element inductors and grAl devices in general. A similar degradation of the internal quality factor in the single-photon regime with increasing sheet resistance has also been observed in some other high kinetic inductance materials like NbN and amorphous InO\,\cite{Frasca_NbN_2023,charpentier2024}. Interestingly, our analysis indicates that our lowest resistivity grAl resonators have surface losses comparable to 3D Al transmons. 

Furthermore, we have demonstrated high quality factors in hybrid resonators that integrate these high-impedance grAl inductors with low-impedance Al or Ta capacitors. Our method of using a dedicated Al bandage layer to make the contact allows grAl to be compatible with a variety of fabrication processes. Since we observe the same trend with resistivity in all devices, we suspect that increasing losses are caused by increasing conductor losses in the grAl strip. These findings motivate future studies that can uncover the microscopic mechanisms that contribute to losses in grAl films and the contacts in hybrid devices. 

Finally, we measure millisecond timescales for QP relaxation in our all-grAl resonators. This is significantly shorter than previous reports on grAl resonators, but comparable to lifetimes measured in Al transmons with similar geometry. Our results merit further investigation into the role of QP localization in grAl films as a function of resistivity and thickness, and how it affects the measured lifetime. Moreover we find that interfacing grAl with a higher-gap superconductor like Ta via an Al bandage still results in millisecond timescales for QP relaxation. These grAl inductors can now be readily integrated into circuits potentially limited by the non-idealities of Josephson junction arrays or geometric inductances.

\begin{acknowledgements}
We thank Ioan M. Pop, Leonid Glazman, Harvey Moseley, Alex Read and Pavel Kurilovich for fruitful discussions. This research was sponsored by the Army Research Office (ARO) under grant no. W911NF-23-1-0051, and by the Air Force Office of Scientific Research under grant no. FA9550-23-1-0338. The views and conclusions contained in this document are those of the authors and should not be interpreted as representing the official policies, either expressed or implied, of the ARO, AFOSR or the US Government. The US Government is authorized to reproduce and distribute reprints for Government purposes notwithstanding any copyright notation herein. Fabrication facilities use was partially supported by the Yale University Cleanroom, a core facility under the directorate of the Provost Office. We thank Yong Sun, Lauren McCabe and Kelly Woods for guidance and assistance towards developing and implementing fabrication processes. We also acknowledge the support of the Yale Quantum Institute. 

L.F. and R.J.S. are founders and shareholders of Quantum Circuits Inc. (QCI).

\end{acknowledgements}
\section{Author Contributions}
P.W. conceptualized the experiment. V.G. developed the fabrication process and fabricated the grAl, grAl/Al samples. N.T. fabricated the grAl/Ta samples. V.G., P.W. and N.T. designed and measured the samples, and analyzed the data. V.G. performed loss analysis with assistance from Y.W. and P.v.V. S.G. designed and prepared the measurement package. L.F. assisted with maintaining the fabrication and experimental capabilities. R.J.S. supervised the project. V.G., P.W., N.T., L.F. and R.J.S. wrote the manuscript with feedback from coauthors. 

\appendix

\section{Contribution of package-losses to the measured internal loss}
\label{appendix_package_loss}
To quantify the contribution of package-losses ($1/Q_{\mathrm{pkg}}$) to the measured internal loss of the resonators $(1/Q_{\mathrm{int}})$, we consider three loss mechanisms: surface dielectric loss at the metal-air (MA) interface of the package, conductor loss in the package and losses at the seams that form between the tunnel and end-caps of the package\cite{ganjam_surpassing_2024}. 
\begin{equation}
\frac{1}{Q_{\mathrm{pkg}}} = 
p_{\mathrm{pkg_{\mathrm{MA}}}}\Gamma_{\mathrm{pkg_{\mathrm{MA}}}}+
p_{\mathrm{pkg_{\mathrm{cond}}}}\Gamma_{\mathrm{pkg_{\mathrm{cond}}}} + 
\frac{y_{\mathrm{seam}}}{g_{\mathrm{seam}}}
\end{equation}
where $\Gamma_{\mathrm{pkg_{\mathrm{MA}}}}$ is the surface dielectric loss tangent of the MA interface, $\Gamma_{\mathrm{pkg_{\mathrm{cond}}}}$ is the package conductor loss factor, $g_{\mathrm{seam}}$ is the seam conductance and $p_{\mathrm{pkg_{\mathrm{MA}}}},p_{\mathrm{pkg_{\mathrm{cond}}}}$ are the corresponding participation ratios and $y_{\mathrm{seam}}$ is the seam admittance. The loss factors for conventionally machined 5N5 Aluminum have been measured in \cite{lei2023characterization} to be $\Gamma_{\mathrm{pkg_{\mathrm{MA}}}} = (4.1\,\pm\,1.8) \times 10^{-2}$, $\Gamma_{\mathrm{pkg_{\mathrm{cond}}}} = R_{\mathrm{s}}/(\mu_0\omega\lambda)$ where $R_{\mathrm{s}} = (0.61\,\pm\,0.28)\,\si{\micro\Omega}$ and $\lambda = 50\,$nm is the effective penetration depth of the superconductor. Further, the tunnel package used in this experiment had been characterized to have a $1/g_\mathrm{seam} = (4.75\,\pm\,4.5) \times 10^{-3}\,\si{\Omega\metre}$ in \cite{ganjam_surpassing_2024}. Using these average values, along with the simulated package participation ratios estimated for our lumped element resonators  we estimate the limit on the package loss to be $Q_{\mathrm{pkg}} \gtrsim 78 \times 10^{6}$. This number is significantly larger than the measured $Q_{\mathrm{int}}$ of our resonators, resulting in a negligible contribution to the total loss. A breakdown of these losses for the resonator that is estimated to have the highest package loss ($f_{\mathrm{r}} =$ 4.60\,GHz, $l_{\mathrm{strip}} =60\,\si{\micro\metre}$, $w_{\mathrm{strip}} =2\, \si{\micro\metre}$, $\rho_{\mathrm{n}}\sim 4160\,\si{\micro\Omega}$cm,  $d = 8\,$mm) is given in Table\,\ref{table_package}.

\renewcommand{\arraystretch}{1.75}
\begin{table}[h]
    \centering
    \caption{\textbf{Summary of package losses.} The different contributions to the package loss are dielectric losses arising in the MA interface of the package surface, the conductor loss, and the seam loss between the main body of the package and the sample clamps. The contributions are estimated for resonator ($f_{\mathrm{r}} =$ 4.60\,GHz, $l_{\mathrm{strip}} =60\,\si{\micro\metre}$, $w_{\mathrm{strip}} =2\, \si{\micro\metre}$, $\rho_{\mathrm{n}}\sim 4160\,\si{\micro\Omega}$cm,  $d = 8\,$mm) that has lowest limit on $Q_{\mathrm{pkg}}$ according to our simulations. }
    \begin{tabular}{|c|c|c|c|}
        \hline
        \textbf{Package} & \textbf{Loss factor} & \textbf{Participation} & \textbf{Loss}  \\
        \hline
         \hline
        MA  & $4.1 \times 10^{-2}$ & $5 \times 10^{-9}$ & $2\times 10^{-10}$ \\
        \hline
       conductor & $2.1\times 10^{-3}$ & $1.7 \times 10^{-6}$ & $3.6 \times 10^{-9}$\\
       \hline
       seam & $4.75 \times10^{-3}\,(\si{\Omega\metre})$ & $1.9 \times 10^{-6}\,(\si{\Omega\metre})^{-1}$  & $9\times 10^{-9}$ \\
       \hline
    \end{tabular}

    \label{table_package}
\end{table}

\section{Device fabrication and sample information} 
\label{appendix_fab}
\begin{figure}[t]

    \includegraphics[width =\linewidth]{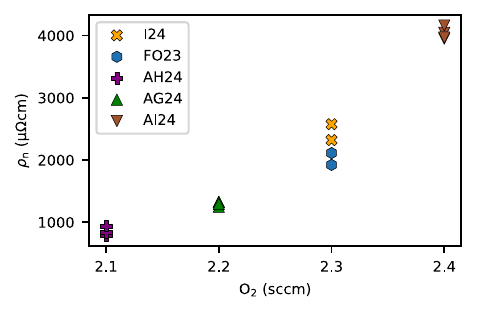}
    \caption{\textbf{Room-temperature resistivity of measured samples vs oxygen flow used rate while evaporating Al}. The corresponding oxygen partial pressure in the chamber was $\sim 5 \times 10^{-5}\,$ mbar. Different colors and symbols correspond to different wafers. These wafers were fabricated using Al from the same crucible without refilling between runs.}
    \label{fig_rho_vs_O2}
\end{figure}

All resonators were fabricated on sapphire substrates grown using the edge-fed film growth (EFG) method and patterned using photolithography. For the all-grAl devices, the wafers were first solvent cleaned by sonication in \textit{N}-Methylpyrrolidone (NMP), acetone and ispropanol, followed by a DI water rinse. The wafers were then dehydrated and primed with hexamethyldisilizane (HMDS) in a YES-310TA Vapor Prime Oven and immediately coated with a bilayer photoresist stack of LOR 5A and S1805. The exposure was done using a Heidelberg MLA 150 Direct Writer. After a 75 sec development in Microposit MF319 developer, the wafers were treated with oxygen plasma using an AutoGlow 200 at 100\,W, 300\,mTorr for 2 minutes. To deposit the grAl film, the wafer was then loaded into a Plassys electron-beam evaporator, where aluminum was evaporated at zero angle while flowing oxygen and rotating the wafer at $\sim25$\,deg/sec. The corresponding oxygen partial pressure in the chamber was $\sim 5 \times 10^{-5}$\,Torr and the Al was evaporated at 1\,nm/sec. The film was lifted off in a hot NMP bath at $\sim 90\,^{\circ}$C for a few hours followed by rinsing in NMP, acetone, isopropanol and DI water. The resistivity of the grAl film was then probed using four-probe van der Pauw measurements.  The measured room-temperature resistivity of the films is plotted vs the corresponding oxygen flow rate used during evaporation in Fig.\,\ref{fig_rho_vs_O2}. 

For the hybrid grAl/Al samples, the grAl strip was fabricated as described above. The 80\,nm thick Al pads were fabricated in a second round of patterning following the same steps. The bandage layer was fabricated in a third round of patterning also following the same procedure but with an additional thin layer ($< 100\,$nm) of 495 PMMA A3 spun before spinning the photoresist stack. The PMMA was used to protect the pre-patterned grAl and Al features from being etched by the MF319 developer. After development, the PMMA layer was removed by oxygen plasma in the AutoGlow 200 at 100\,W, 300\,mtorr for 3 minutes. Additionally, an in-situ Ar ion beam milling at 400\,V, 15\,mA, 4\,sccm was done for 4 min to remove the oxide layers prior to the Al bandage deposition.

For the hybrid grAl/Ta samples, the EFG sapphire substrates were first cleaned in a piranha solution (2:1 H$_2$SO$_4$:H$_2$O$_2$) and annealed at 1200\,$^{\circ}$C in an oxygen-rich environment. Then 75\,nm of Ta was deposited on the entire wafer by DC magnetron sputtering at an elevated temperature of 800\,$^{\circ}$C using an Ar pressure of 6\,mTorr at 2.5\,$\mathrm{\AA}$/sec. The Ta film was then coated with S1827 and the photoresist was patterned using photolithography, followed by development in MF319. The wafer was then hard-baked for 1 min at 120\,$^{\circ}$C and oxygen plasma cleaned in the AutoGlow 200 at 150\,W, 300\,mTorr for 2 mins to remove resist residue. The Ta was then etched at a rate of 100\,nm/min in an Oxford 80+ Reactive Ion Etcher using SF$_6$ with a flow rate of 20\,sccm, a pressure of 10\,mTorr, and an RF power of 50\,W. After etching, the wafer was sonicated in NMP, acetone, isopropanol and DI water to remove the photoresist. The wafer was then piranha cleaned again and the Ta oxide was stripped using Transene 10:1 BOE for 20 mins. This was followed by patterning of the grAl strip and Al bandage layers using the procedure described earlier. 

After device fabrication, all wafers were coated with protective resist and diced using an ADT ProVectus 7100 dicer followed by a cleaning in NMP, acetone, isopropanol and DI water.  A summary of the grAl film parameters, resonator dimensions, measured frequencies and quality factors of all the samples is given in Table \ref{table_info}.

\section{Measurement setup} 
\label{appendix_setup}
 
\begin{figure}[t]
    \includegraphics[width =\linewidth]{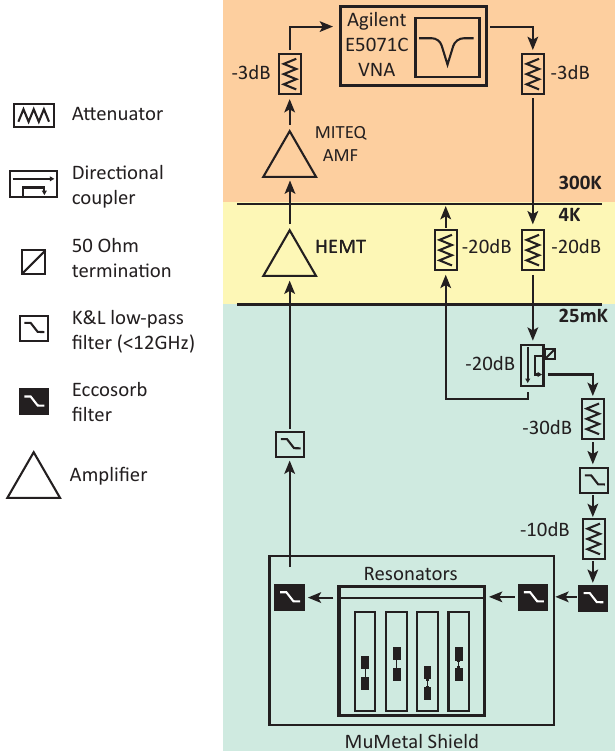}
    \caption{\textbf{Fridge wiring diagram}. Experimental setup for measuring resonator transmission coefficient using a VNA. 
    \label{fig_fridge_wiring}
    }
\end{figure}

The RF wiring setup used for characterizing the resonators is shown in Fig.\,\ref{fig_fridge_wiring}. The resonators are housed in an Al package which is mounted to the mixing chamber stage of a dilution refrigerator. The input and output ports of the package use custom CR-110 IR Eccosorb low-pass filters to reduce the exposure of our samples to pair-breaking infrared radiation. The package and Eccosorb filters are placed within a MuMetal can to screen external magnetic fields. Both, input and output lines, are filtered with 12 GHz K\&L low-pass filters.

Across all stages, the input line has approximately 90\,dB attenuation in the 4-8 GHz range based on room temperature calibration. There is 20\,dB of attenuation at the 4K stage. The remaining 70\,dB of attenuation is at the mixing chamber stage, with 20\,dB of that attenuation achieved by a directional coupler similar to Ref.\,\cite{ganjam_surpassing_2024}.

The first amplifier stage is a high-electron-mobility transistor (HEMT) amplifier anchored at the 4 K stage. At room temperature, there is an additional MITEQ amplifier before the output signal is detected by a vector network analyzer (VNA). All frequency and time-domain measurements were performed using an Agilent E5071C VNA which was clocked to an SRS FS725 10\,MHz Rb frequency standard.

\section{Explanation for corrected Kerr with strip length}
\label{App_Kerr_correction}
In Sec.\,\ref{SEC_Lkin_Kerr} of the main text we calculate the number of effective junctions $N_{\mathrm{JJ}}$ and the critical current density $J_{\mathrm{c}}$ from the measured self-Kerr coefficient $K$ of the all-grAl samples. Since we are mostly interested in the parameters of the grAl strip, we need to account for the finite reduction of the measured self-Kerr coefficient due to the presence of additional inductance in the leads, which connect the strip to the capacitor electrodes (see Fig.\,\ref{Fig_sample_design}). Since we keep a minimum distance of $300\,\si{\micro\metre}$ between the capacitor electrodes, the leads become increasingly long with decreasing strip length. Hence, especially for the highest resistivity all-grAl samples with the shortest strip length, we need to account for the additional kinetic inductance in the leads and the capacitor electrodes. We can estimate the reduction in the measured self-Kerr coefficient by calculating the energy participation ratio $p = L_{\mathrm{strip}} / L_{\mathrm{tot}}$ of the strip with respect to the total inductance in the circuit\,\cite{Minev2021}. We find $K = p^2 K_{\mathrm{strip}}$, where $K_{\mathrm{strip}} = E_{\mathrm{c}} / N_{\mathrm{JJ}}^2$ is the non-linearity we would measure if the strip would be the only inductive element. For the samples shown in Fig.\,\ref{Fig_Lkin}, the inductive contribution of the leads and the capacitor electrodes are visible as the $y$-axis offset. Additionally, we also account for small variations in the single electron charging energy $E_{\mathrm{c}}$ with changes in the sample geometry by simulating each configuration individually in HFSS. 

\section{Two-level system model}
\label{TLS_model}
The power dependence of the measured $Q_{\mathrm{int}}$ shown in Fig.\,\ref{Fig_Qs} is fit to a general model that describes a TLS environment\,\cite{Gao_2008,Pappas_2011_TLS} or any other mechanism that saturates with power: 
\begin{equation}
    \frac{1}{Q_{\mathrm{int}}(\overline{n}_{\mathrm{ph}})} = \frac{1}{Q_{\mathrm{0}}} 
    +\frac{p_{\mathrm{surf}}\tan\delta_{\mathrm{TLS}}}{\sqrt{1+(\overline{n}_{\mathrm{ph}}/n_{\mathrm{c}})^{\beta}}}
\end{equation}
where $1/Q_{\mathrm{0}}$ is the power-independent contribution to the total internal loss, $\tan\delta_{\mathrm{TLS}}$ is the ensemble TLS loss tangent, $n_{\mathrm{c}}$ is the critical photon number for saturation and $\beta$ describes TLS interaction. As described in Sec.\,\ref{sec_loss_all_grAl}, we use this fit only as an interpolating function to extract the internal quality factor in the single photon regime.
\section{Power dependence of residual loss}

\begin{figure}[t]
    \includegraphics[width =0.85\linewidth]{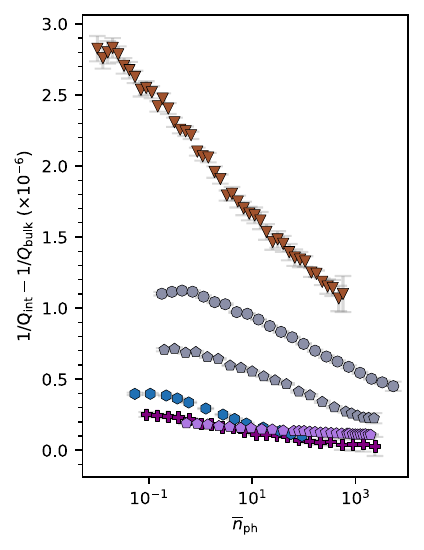}
    \caption{\textbf{Power dependence of residual loss for six representative samples (see Tab.\,\ref{table_info}}).}
    \label{fig_loss_vs_nbar}
\end{figure}

The power dependence of the residual loss for six representative samples spanning the entire resistivity range is shown in Fig.\,\ref{fig_loss_vs_nbar}. Independent of the resistivity or type (all-grAl and hybrid), all samples show a power dependence that can be described by a generic saturation model (App.\,\ref{TLS_model}). Interestingly, since the hybrid samples are not limited by surface losses, the observed power dependence suggests that the environment causing the conductor loss is also saturable.

\section{Temporal fluctuations in $Q_{\mathrm{int}}$} \label{appendix_Q_fluct}
\begin{figure}[h]
\includegraphics[width =\linewidth]{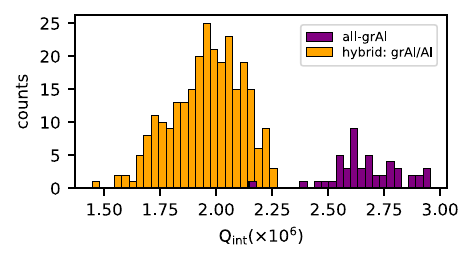}
    \caption{\textbf{Temporal fluctuations of low-power $Q_{\mathrm{int}}$:} Histogram of fluctuations in the internal quality factor for an all-grAl resonator ($f_{\mathrm{r}} =$ 4.79\,GHz, $l_{\mathrm{strip}} =480\,\si{\micro\metre}$, $w_{\mathrm{strip}} =2\, \si{\micro\metre}$, $\rho_{\mathrm{n}}\sim 830\,\si{\micro\Omega}$cm) measured over 32 hours with $\overline{n}_{\mathrm{ph}} = 1$ and for a hybrid Al/grAl resonator ($f_{\mathrm{r}} =$ 7.29\,GHz, $l_{\mathrm{strip}} =294\,\si{\micro\metre}$, $w_{\mathrm{strip}} =10\, \si{\micro\metre}$, $\rho_{\mathrm{n}}\sim 2250\,\si{\micro\Omega}$cm) measured over 180 hours at a low power with $\overline{n}_{\mathrm{ph}} = 0.5$.}
    \label{fig_Q_fluctuations}
\end{figure}

The internal quality factor of the resonators was monitored over several days at low input drive powers close to the single-photon regime. A histogram of the measured $Q_{\mathrm{int}}$ for two representative samples: an all-grAl resonator ($f_{\mathrm{r}} =$ 4.79\,GHz, $l_{\mathrm{strip}} =480\,\si{\micro\metre}$, $w_{\mathrm{strip}} =2\, \si{\micro\metre}$, $\rho_{\mathrm{n}}\sim 830\,\si{\micro\Omega}$cm) measured over 32 hours and for a hybrid grAl/Al sample ($f_{\mathrm{r}} =$ 7.29\,GHz, $l_{\mathrm{strip}} =294\,\si{\micro\metre}$, $w_{\mathrm{strip}} =10\, \si{\micro\metre}$, $\rho_{\mathrm{n}}\sim 2250\,\si{\micro\Omega}$cm) measured over 180 hours is shown in Fig. \ref{fig_Q_fluctuations}. We find that the quality factors are quite stable with fluctuations of $< \pm10\%$ about the mean value over these long timescales. These fluctuation could arise due to fluctuating TLSs in the surface or bulk dielectrics. 

\section{Surface loss factor of pure aluminum}
\label{appendix_surface_loss_Al}
Our loss characterization uses the $\Gamma_{\mathrm{surf}}$ for Al from \cite{ganjam_surpassing_2024}, which depends on both the material and its corresponding fabrication processes. However, the Al devices in \cite{ganjam_surpassing_2024} are made entirely with electron-beam  lithography while our resonators are patterned entirely with photolithography. To determine whether the use of photolithography significantly alters the $\Gamma_{\mathrm{surf}}$ for Al, we build an Al tripole stripline resonator similar to \cite{ganjam_surpassing_2024} and characterize the three lowest frequency modes. We find the measured internal quality factors of the different modes of our Al tripole to be consistent with the predictions from \cite{ganjam_surpassing_2024}, indicating that the $\Gamma_{\mathrm{surf}}$ for Al does not differ significantly between the two processes. 

\section{Procedure for analyzing high-energy impacts in resonators}
\label{appendix_impacts_procedure}
\begin{figure}[t]
    \includegraphics[width =\linewidth]{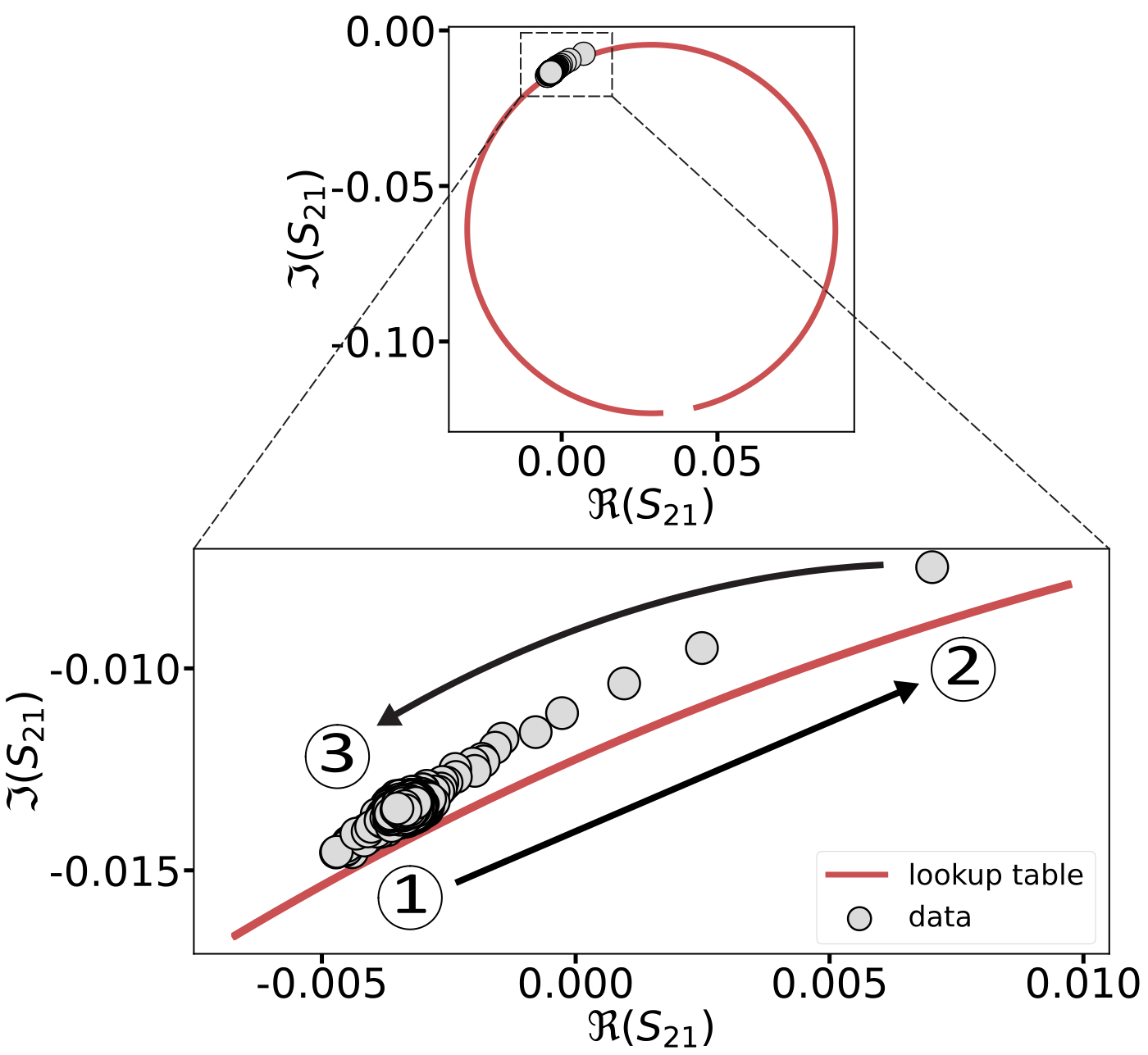}
    \caption{\textbf{Complex response of an all-grAl resonator during a high energy impact:}  The time trace data (grey markers) for the all-grAl resonator from Fig. \,\ref{Fig_impacts} ($f_{\mathrm{r}} =$ 4.60\,GHz, $l_{\mathrm{strip}} =60\,\si{\micro\metre}$, $w_{\mathrm{strip}} =2\, \si{\micro\metre}$, $\rho_{\mathrm{n}}\sim 4160\,\si{\micro\Omega}$cm, $Q_{\mathrm{c}}\sim 400$) is plotted on top of its corresponding resonance circle (red). The data was measured using an IF bandwidth = 50\,kHz and an average photon number $\bar{n} \sim 10^5$. The numbered circles and black arrows indicate the chronological order of the points in the time trace. Immediately before the high energy impact (point 1), the resonator is in equilibrium and the transmission coefficient is near the corresponding on-resonance point. Immediately after an impact (point 2), the increase in non-equilibrium QPs causes a decrease in the resonance frequency. After some time, the resonator moves back towards the initial on-resonance point (point 3). Thus, the circle can be used as a lookup table to convert the complex transmission coefficients from the time trace to a shift in resonance frequency \cite{guenzler2021}.}
    \label{fig_grAl_impact_circle}
\end{figure}

\begin{figure}[t]
    \includegraphics[width =\linewidth]{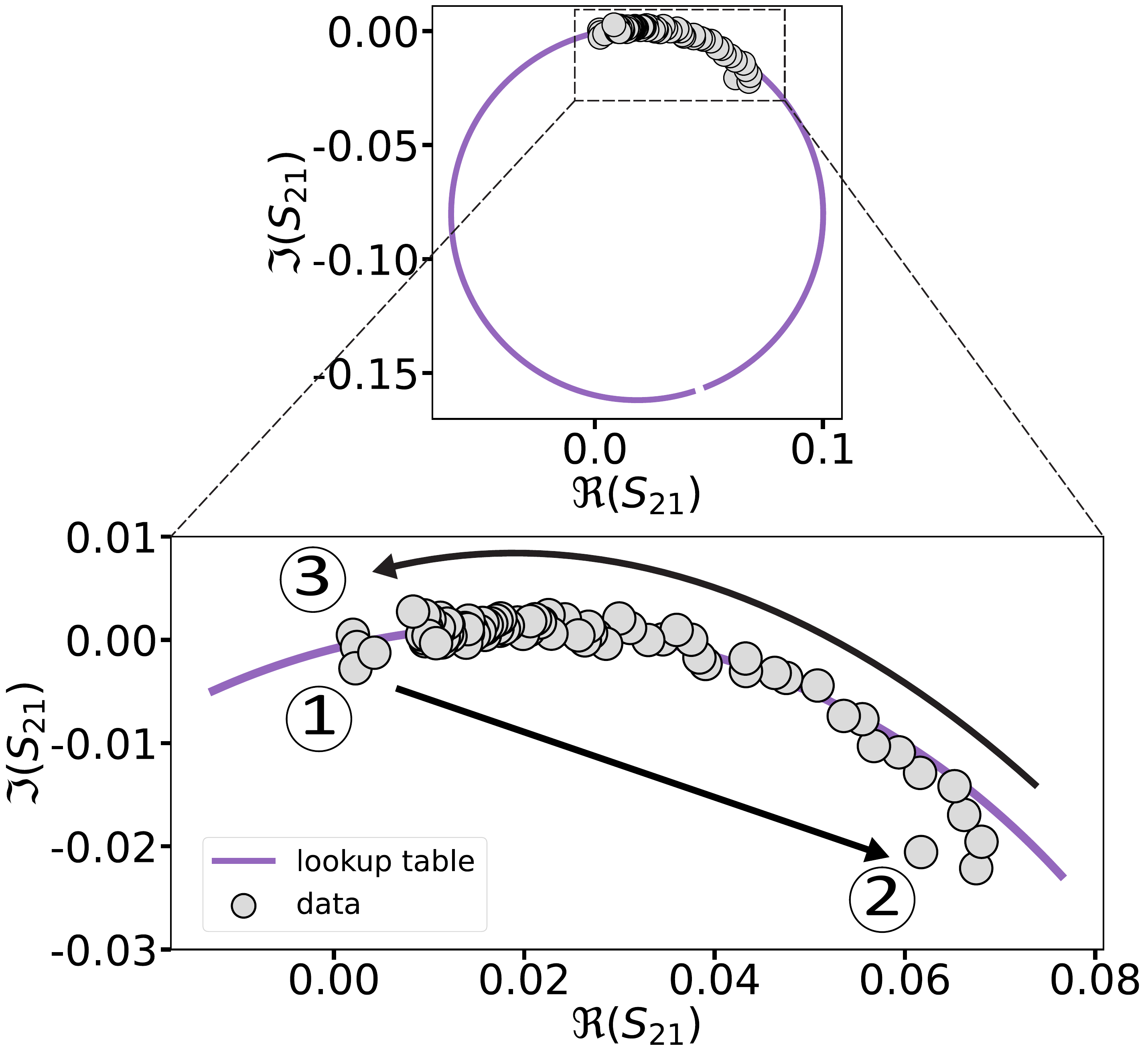}
    \caption{\textbf{Complex response of a grAl/Ta resonator during a high energy impact:} A grAl/Ta resonator’s ($f_{\mathrm{r}} =$ 5.19 GHz, $l_{\mathrm{strip}} =$  200 $\si{\micro\metre}$, $w_{\mathrm{strip}} = $ 5 $\si{\micro\metre}$ , $\rho_{\mathrm{n}}=$ 2700 $\si{\micro\Omega}$cm) transmission coefficient was measured (at IF bandwidth = 50 kHz and $\bar{n} \sim 10^4$) as a function of time (grey markers) during a high energy impact. It is plotted on top of a resonance circle (purple) which was separately obtained by measuring (at $\bar{n} \sim 10^4$) the resonator’s transmission coefficient as a function of drive detuning and fitting the response.}
    \label{fig_hybrid_impact_circle}
\end{figure}

To detect the signature of high energy impacts into the sapphire substrate in the vicinity of our resonators, we monitor the complex transmission coefficient $S_{\mathrm{21}}$ close to resonance as a function of time by operating a VNA in zero-span mode at a fixed readout power. We separately measure the transmission coefficient of the resonator as a function of drive detuning at the same power as the time trace measurement. The resonator response forms a circle when plotted in the complex plane where each point on the circle corresponds to a unique detuning relative to the resonance frequency. 

The time trace for the all-grAl sample shown in Fig.\,\ref{Fig_impacts} of the main text, taken during a high energy impact, is plotted on top of its corresponding circle in Fig.\,\ref{fig_grAl_impact_circle}. The numbers indicate the chronological order of the time trace data. Prior to the impact (point 1), the response is near the on-resonant point on the circle. Immediately after an impact, the resonator’s frequency decreases, and the complex response moves away from the on-resonant point (point 2). Eventually, the resonator relaxes back to its equilibrium frequency (point 3) and the complex response traverses the circle. {Since the location in the complex plane depends only on the relative detuning, the circle can be used as a lookup table to map the complex response from the time trace to a resonator frequency shift $f(t)$\,\cite{guenzler2021}.

Further, the extracted frequency shift $\delta f(t)$ can be related to a change in the quasiparticle density as \cite{valenti_interplay_2019}
\begin{equation}
    \frac{\delta f(t)}{f} = -\frac{\alpha}{4}\delta x_{\mathrm{qp}}(t).
\end{equation}

To fit the time-dependence of $\delta x_{\mathrm{qp}}(t)$ using the phenomenological model described in the main text, we use the following analytical solution as described in\,\cite{Wang2014}.
\begin{equation}
    \delta x_{\mathrm{qp}}(t) = x_{\mathrm{qp}} - x^0_{\mathrm{qp}} = x_{\mathrm{i}} \frac{1-r^{\prime}}{e^{t/\tau_{\mathrm{ss}}}-r^{\prime}}
    \label{eq_xqp_fit}
\end{equation}
where $\tau_{\mathrm{ss}} = 1/(2rx^0_{\mathrm{qp}}+s)$ is the time scale of the exponential decay, $x_{\mathrm{i}}$ is the initial QP density right after the burst, and $x^0_{\mathrm{qp}}$ is the steady state QP density which is estimated from the residual loss as\,\cite{Martinis_2009_QPs, barends2011minimizing} 
\begin{equation}
    x^0_{\mathrm{qp}} = \frac{\pi}{\alpha Q_{\mathrm{res}}} \sqrt{\frac{\hbar \omega}{2\Delta}}.
\end{equation}
Using the parameters obtained by fitting Eq.\,\ref{eq_xqp_fit} to the data, we calculate the recombination constant, trapping rate and background generation rate as 
\begin{equation}
    r = \frac{r^{\prime}}{(1-r^{\prime})\tau_{\mathrm{ss}}x_{\mathrm{i}}} 
\end{equation}
\begin{equation}
    s = \frac{1}{\tau_{\mathrm{ss}}}\left[1 - \frac{2r^{\prime}x^0_{\mathrm{qp}}}{(1-r^{\prime})x_{\mathrm{i}}}\right]
\end{equation}
\begin{equation}
    g = \frac{x^0_{\mathrm{qp}}}{\tau_{\mathrm{ss}}}\left[1 - \frac{r^{\prime}x^0_{\mathrm{qp}}}{(1-r^{\prime})x_{\mathrm{i}}}\right]
\end{equation}
Since we only use an upper bound for the steady state QP density, there is additional uncertainty when estimating the model parameters $r$, $s$, and $g$.

\section{Critical temperature}
To extract the critical temperature $T_{\mathrm{c}}$ of our grAl films, we monitor the response of an all-grAl resonator as a function of (fridge) temperature. In general, as the sample temperature is increased and approaches $T_\mathrm{c}$, the number of thermal quasiparticles increases resulting in a downward shift of the resonator frequency and an increase in conductor loss.  This shift in frequency $\Delta f$ and change in quality factor $\Delta(1/Q_{\mathrm{cond}})$ with temperature can be related to a change in the complex surface impedance $Z_\mathrm{s} = R_{\mathrm{s}}+iX_{\mathrm{s}}$ of the superconductor as \cite{Gao_2008} 
\begin{equation}
    \frac{f(T)-f(0)}{f(0)} = -\frac{\alpha}{2}\frac{X_\mathrm{s}(T)- X_\mathrm{s}(0)}{X_\mathrm{s}(0)}
\end{equation}
\begin{equation}
    \frac{1}{Q_{\mathrm{cond}}(T)} - \frac{1}{Q_{\mathrm{cond}}(0)} = \alpha \frac{R_{\mathrm{s}}(T) - R_{\mathrm{s}}(0)}{X_\mathrm{s}(0)}
\end{equation}
This change in the surface impedance is a consequence of the change in the complex conductivity $\sigma$ of the superconducting film. For a thin film where $t \ll \lambda$, the two quantities and can be related as: 
\begin{equation}
\frac{\delta Z_\mathrm{s}(T)}{Z_{\mathrm{s}}(0)} = - \frac{\delta\sigma(T)}{\sigma(0)}
\end{equation}
Further, the temperature dependence of $\sigma$ can be calculated by numerical integration of the Mattis-Bardeen equations for a BCS superconductor\,\cite{mattis1958theory, Gao_2008, krayzman2022thin}. We use this model to fit the temperature dependence of the resonator response and extract $T_{\mathrm{c}}$ as the only fit parameter while assuming a fixed $\alpha$.  Here, we study the temperature dependence of an all-grAl resonator ($f_{\mathrm{r}} =$ 4.60\,GHz, $l_{\mathrm{strip}} =60\,\si{\micro\metre}$, $w_{\mathrm{strip}} =2\, \si{\micro\metre}$, $\rho_{\mathrm{n}}\sim 4160\,\si{\micro\Omega}$cm, $Q_{\mathrm{c}}\sim 400$,  $\alpha=0.96$). The resonator is deliberately over-coupled to the input port to track $\Delta f$ to higher temperatures closer to $T_{\mathrm{c}}$. Consequently, we only fit to the shift in resonator frequency with temperature as shown in Fig.\,\ref{MB_fvsT}. We extract $T_{\mathrm{c}} = 2.15\,$K for the grAl film which is consistent with previous reports of the critical temperature of grAl in this resistivity range \cite{levy2019electrodynamics}.

\begin{figure}[t]
\includegraphics[width =\linewidth]{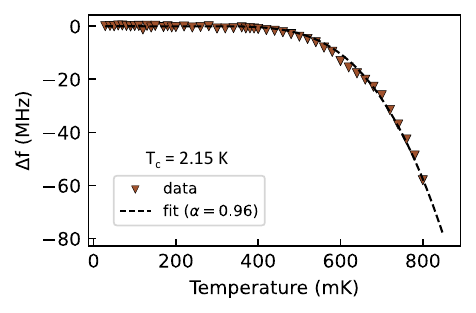}
    \caption{\textbf{Temperature dependence of resonance frequency}: The shift in resonance frequency $\Delta f$ of an all-grAl resonator ($f_{\mathrm{r}} =$ 4.60\,GHz, $l_{\mathrm{strip}} =60\,\si{\micro\metre}$, $w_{\mathrm{strip}} =2\, \si{\micro\metre}$, $\rho_{\mathrm{n}}\sim 4160\,\si{\micro\Omega}$cm, $Q_{\mathrm{c}}\sim 400$), is plotted as a function of temperature. The frequency shift is related to a change in surface impedance of the superconductor and calculated using the Mattis-Bardeen equations for a BCS superconductor. The solid line is a fit to the data obtained from numerical integration of the surface impedance assuming a fixed kinetic inductance fraction $\alpha=0.96$, resulting in a critical temperature $T_{\mathrm{c}} = 2.15\,$K for the grAl film.}
    \label{MB_fvsT}
\end{figure}

\section{Impedance of inductor strip in a coaxial tunnel}
In this study, we do not focus on building high impedance circuits. However, it is still useful to estimate the characteristic impedance $Z_0$ of our grAl inductors in the coaxial tunnel geometry. Assuming an outer tunnel diameter of $4\,\mathrm{mm}$, we calculate the characteristic impedance of the inductors from the capacitance per unit length $c_0$ and the kinetic inductance per unit length $l_{\mathrm{kin}} = L_{\mathrm{sq}}/w$. The capacitance to ground increases only weakly when increasing the width of the strip from $2\,\si{\micro\metre}$ to $10\,\si{\micro\metre}$: $c_0 = 6.6 - 8.4\,\mathrm{aF}/\si{\micro\metre}$. Hence, for our highest resistivity films, we find a characteristic impedance of $Z_0 = 4.92\,\si{\kilo\ohm}$ for the narrowest strip width. In order to increase this number well beyond the resistance quantum $R_\mathrm{q} \simeq 6.48\,\si{\kilo\ohm}$, we would need to either narrow down the strip more, or reduce the thickness of the film.

\renewcommand{\arraystretch}{1.75}

\begin{table*}[t!]
\centering
\caption{\textbf{Summary of resonator parameters, measured frequencies and quality factors} The wafer label indicates the different batches used in our study, while the device type describes the materials used. The other resonator parameters are the sheet resistance $R_{\mathrm{sq}}$ measured at room temperature, the normal-state resistivity $\rho_{\mathrm{n}} = R_{\mathrm{s}} t$ calculated from the sheet resistance and the film thickness, as well as the length and the width of the strip $l_{\mathrm{strip}}$ and $w_{\mathrm{strip}}$, respectively. The kinetic inductance fraction $\alpha = 1 - L_{\mathrm{g}}C_{\mathrm{s}}\omega_{\mathrm{r}}^2$ is estimated from the measured resonance frequency $f_{\mathrm{r}}$ and the simulated geometric inductance $L_{\mathrm{g}}$ and shunt capacitance $C_{\mathrm{s}}$. The internal quality factor $Q_{\mathrm{int}}$ is given in the single-photon regime for all samples that have a reasonable ratio of internal to coupling quality factor. }
\begin{tabular}{|c|c|c c|c c c|c|c c|c|c|}
\hline

\textbf{Wafer} &
\textbf{Type} &
\textbf{$R_{\mathrm{sq}}$} &
\textbf{$\rho_{\mathrm{n}}$} & 
\textbf{$l_{\mathrm{strip}}$} &
\textbf{ $w_{\mathrm{strip}}$} & 
\textbf{$f_{\mathrm{r}}$} & 
\textbf{$\alpha$} &
\textbf{$Q_{\mathrm{int}}$ } &
\textbf{$Q_{\mathrm{c}}$ }  &
\textbf{marker}\\

 & & $(\Omega)$ & $(\si{\micro\Omega}$cm) & $(\si{\micro\metre})$ & $(\si{\micro\metre})$ & (GHz) & & $(\times 10^6)$ & $(\times 10^6)$ & \\
\hline 
\hline
AH24  & all-grAl & 91 & 827 & 480 & 2 & 4.79 & 0.93 & 2.67 & 1.65 &\includegraphics[height=0.25cm]{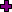}\\

AH24  & all-grAl & 86 & 786 & 400 & 2 & 5.23 & 0.93 & 2.30 & 1.48 & \includegraphics[height = 0.25cm]{symbols/AH24.pdf} \\

AH24 &all-grAl & 92 & 839 & 320 & 2 & 5.67 & 0.93 & 1.83 & 1.60 &  \includegraphics[height = 0.25cm]{symbols/AH24.pdf} \\

AH24  &all-grAl & 103  & 938 & 240 & 2 & 6.32 & 0.91 & 2.38 & 2.15 &\includegraphics[height = 0.25cm]{symbols/AH24.pdf} \\
\hline
\hline
AG24  &all-grAl & 142 & 1293 & 240 & 2 & 5.27 & 0.94 & 1.67 & 1.97 &\includegraphics[height = 0.25cm]{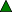}\\

AG24  &all-grAl & 137 & 1248 & 200 & 2 & 5.69 & 0.94 & 1.59 & 2.73& \includegraphics[height = 0.25cm]{symbols/AG24.pdf}\\

AG24  &all-grAl & 145 & 1323 & 160 & 2 & 6.16 & 0.93 &  1.61 & 0.93 & \includegraphics[height = 0.25cm]{symbols/AG24.pdf}\\

AG24   &all-grAl & 146 &1329& 120 & 2 & 6.78 & 0.92 &  1.33 & 1.62 & \includegraphics[height = 0.25cm]{symbols/AG24.pdf}\\
\hline
\hline
FO23  &all-grAl & 232 & 2111 & 150 & 3 & 6.04 & 0.94 & 1.78 & 1.30 & \includegraphics[height = 0.25cm]{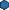} \\

FO23   &all-grAl & 211 & 1920 & 400 & 10 & 6.72 & 0.90 & 1.97 & 1.95 & \includegraphics[height = 0.25cm]{symbols/FO23.pdf}\\
\hline
\hline
I24  & all-grAl & 255 & 2320 & 400 & 10 & 6.13 & 0.92 & 1.89 & 1.38 & \includegraphics[height = 0.25cm]{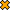} \\

I24   &all-grAl & 283 & 2575 & 500 & 10 & 5.32 & 0.92 & 1.45 & 1.08 & \includegraphics[height = 0.25cm]{symbols/I24_grAl.pdf} \\
\hline
\hline
AI24  & all-grAl & 457 & 4162 & 60 & 2 & 4.60 & 0.96 & 0.52 & 0.68 & \includegraphics[height = 0.25cm]{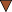}\\

AI24  &all-grAl & 445 & 4045 & 50 & 2 & 4.92 & 0.96 & 0.56 & 0.98 & \includegraphics[height = 0.25cm]{symbols/AI24.pdf} \\

AI24  & all-grAl & 434 & 3952 & 40 & 2 & 5.31 & 0.96 & 0.58 & 0.43 & \includegraphics[height = 0.25cm]{symbols/AI24.pdf} \\

AI24  & all-grAl & 436 & 3966 & 30 & 2 & 5.67 & 0.95 & 0.46 & 0.47& \includegraphics[height = 0.25cm]{symbols/AI24.pdf}\\
\hline
\hline
I24 & hybrid: grAl/Al & 237 & 2155 & 200 & 3.3 & 5.70 & 0.94 & 2.23 & 1.99 & \includegraphics[height = 0.25cm]{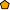} \\

I24 & hybrid: grAl/Al & 247 & 2247 & 294 & 10 & 7.29 & 0.9 & 2.11 & 1.71 & \includegraphics[height = 0.25cm]{symbols/I24_hybrid.pdf} \\
\hline
\hline
GH23 & hybrid: grAl/Ta & 290 & 2639 & 400 & 8 & 4.85 & 0.95 & 1.50 & 0.69 & \includegraphics[height = 0.25cm]{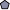} \\

GH23 & hybrid: grAl/Ta & 320 & 2912 & 294 & 10 & 5.97 & 0.93 & 1.33 & 1.30 & \includegraphics[height = 0.25cm]{symbols/GH23_1_hybrid.pdf} \\

GH23 & all-grAl & 300 & 2730 & 400 & 10 & 4.88 & 0.95 & 0.90 & 0.80 & \includegraphics[height = 0.25cm]{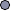} \\

GH23 & hybrid: grAl/Ta & 340 & 3094 & 200 & 5 & 5.19 & 0.95 & $(>> Q_{\mathrm{c}})$ & 0.01 & \includegraphics[height = 0.25cm]{symbols/GH23_1_hybrid.pdf} \\

\hline
\hline

GH23 & hybrid: grAl/Ta & 64 & 582 & 480 & 2 & 5.61 & 0.90 &  4.72 & 1.77 & 
\includegraphics[height = 0.25cm]{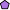} \\

GH23 & all-grAl & 61 & 555 & 400 & 2 & 6.03 & 0.90 &  3.66 & 0.57 & 
\includegraphics[height = 0.25cm]{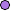} \\

GH23 & hybrid: grAl/Ta & 62 & 564 & 360 & 2 & 6.26 & 0.90 & 3.52 & 6.98 & 
\includegraphics[height = 0.25cm]{symbols/GH23_hybrid.pdf} \\

GH23 & hybrid: grAl/Ta & 66 & 600 & 240 & 2 & 7.00 & 0.90 &  2.37 & 1.06 & 
\includegraphics[height = 0.25cm]{symbols/GH23_hybrid.pdf} \\

\hline
\end{tabular}  
\label{table_info}
\end{table*}

\clearpage

\bibliography{grAl_lumped-element-inductor_refs.bib}
\end{document}